\documentclass[aps,prb]{article}
\usepackage{graphicx,xcolor,soul}
\usepackage{epstopdf}
\usepackage{subfigure}
\usepackage{longtable}

\usepackage[numbers,sort&compress,merge]{natbib}
\bibpunct[, ]{[}{]}{,}{n}{,}{,}

\usepackage[T1]{fontenc}
\usepackage{lmodern}
\usepackage{authblk}
\providecommand{\keywords}[1]{\textbf{\textit{Keywords}} #1}

\begin{document}

\title{Modeling Molecular Magnets with Large Exchange and On-Site Anisotropies}

\author[a]{Sumit Haldar\thanks{E-mail: sumithaldar@iisc.ac.in}}
\author[b]{Rajamani Raghunathan\thanks{E-mail: rajamani@csr.res.in}}
\author[c]{Jean-Pascal Sutter\thanks{E-mail: jean-pascal.sutter@lcc-toulouse.fr}} 
\author[a]{S. Ramasesha\thanks{E-mail: ramasesh@iisc.ac.in}}
\affil[a]{Solid State and Structural Chemistry Unit, Indian Institute of Science, Bangalore - 560012, India.}
\affil[b]{UGC-DAE Consortium for Scientific Research, Indore - 452017, India.}
\affil[c]{LCC-CNRS, Universit\'e de Toulouse, UPS, INPT, CNRS, Toulouse, France.}

\maketitle

\begin{abstract}
\noindent Spins in molecular magnets can experience both anisotropic exchange interactions and on-site magnetic anisotropy. In this paper we study the effect of exchange anisotropy on the molecular magnetic anisotropy both with and without on-site anisotropy. When both the anisotropies are small, we find that the axial anisotropy parameter $D_M$ in the effective spin Hamiltonian is the sum of the individual contributions due to exchange and on-site anisotropies. We find that even for axial anisotropy of about $15\%$, the low energy spectrum does not correspond to a single parent spin manifold but has intruders states arising from other parent spin. In this case, the low energy spectrum can not be described by an effective Hamiltonian spanning the parent spin space. We study the magnetic susceptibility, specific heat as a function of temperature and magnetization as a function of applied field to characterize the system in this limit. We find that there is synergy between the two anisotropies, particularly for large systems with higher site spins.  
\end{abstract}



\baselineskip=22pt
\section{\label{sec:introduction}Introduction}
Molecular spin clusters such as single molecule magnets (SMMs) and single chain magnets (SCMs) have been studied extensively over the last few decades \cite{Sessoli1993, Gatteschi2003, Takahashi1994, Thomas1996, Sessoli1993a, Christou2000}. These spin clusters have attracted huge interest from both theoretical and experimental stand points because of the promise they hold for applications such as in memory storage devices, in quantum computations and in information technologies in general \cite{Tejada2001, Winpenny2008, Lehmann2009, Vincent2012, Troiani2011, Leuenberger2001}. The main bottleneck for these applications appears to be the fast relaxation of the magnetization from the fully magnetized to the non magnetized state. This is due to the low blocking temperature, measured as the temperature at which the relaxation time for magnetization, $\tau_R$, is 100s and depends on the energy barrier between two fully and oppositely magnetized states, for the presently known SMMs and SCMs \cite{Gatteschi2007, Demir2017}. Current research in this field is focused on enhancing the blocking temperature \cite{Woodruff2013, Langley2016}.

The energy barrier $\Delta$, between two fully and oppositely magnetized states of an anisotropic spin cluster of spin $S$ is given by $\Delta=|D_M|S^2$ for an integer spin cluster and $|D_M|(S^2-1/4)$ for a half-integer spin cluster. Therefore, there are two routes to enhancing $\Delta$, $(i)$ by increasing $D_M$ and $(ii)$ by increasing $S$. Increasing $D_M$ can be achieved by using magnetic building blocks in unusual coordination number and geometry. Indeed this has been demonstrated for hepta coordinated complexes \cite{Ruamps2013, Gogoi2013, Venkatakrishnan2010, Bar2015, Bar2017}. Increasing $S$ can be achieved by using rare earth ions in the high spin state as the building blocks. However, it has been shown by Waldmann \cite{Waldmann2007} that the magnetic anisotropy of a ferromagnetic assembly of spins is smaller than the anisotropy of individual spins as each spin center with spin $s_i$ only contributes a fraction 
\begin{eqnarray}
\frac{s_i(2s_i-1)}{S(2S-1)}
\end{eqnarray}
of the site anisotropy to the anisotropy of the SMM or SCM with total spin is $S$. This result assumes that all the individual magnetic ions have non zero axial anisotropy $d_i$ and zero planar anisotropy $e_i$, and that all the spin centers have the same magnetic axes. Notwithstanding this nuance, the result is illustrative of the fact that the anisotropy of the clusters is smaller than that of individual ions.

With $3d$ transition metal complexes, the highest blocking temperature reported is $4.5K$, although the energy barrier $\Delta$ is $62 cm^{-1}$ \cite{Milios2007}. This could be due to the large off-diagonal anisotropy terms that lead to quantum tunneling of magnetization. The anisotropy can be enhanced by choosing ions of $4d$, $5d$ or $4f$ metals wherein the relativistic effects are large, leading to large spin-orbit interactions \cite{Tang2006, Woodruff2013, Rinehart2011, Dreiser2013a, Mironov2003, Bennett2003}. For example, in the $Dy_4$ systems, the energy barrier is $~692 cm^{-1}$ \cite{Blagg2013}. However, large quantum tunneling of magnetization leads to small hysteresis loops. In our previous studies \cite{Haldar2017}, we have shown that large magnetic anisotropy of building blocks leads to breaking the spin symmetry. In this event associating a parent spin state to define the $D_M$ and $E_M$ parameters of a cluster is not possible due to intrusion of states from different parent spins within the given spin manifold. In these cases, the Waldmann conclusion that the contribution of the individual anisotropies decreases with increasing total spin of the cluster is no longer valid. The properties of the system will have to be computed from the eigenstates of the full Hamiltonian.

The origin of single ion anisotropy as well as anisotropic exchange interactions lie in spin-orbit interactions. Indeed, it is difficult to assume isotropic or simple Heisenberg exchange interactions between spin sites that are highly anisotropic. High nuclearity complexes with large anisotropic interactions are known in a few cases, $[Mn^{III}_6Os^{III}]^{3+}$ cluster has $J_x=-9cm^{-1}, J_y=+17cm^{-1}$ and $J_z=-16.5cm^{-1}$ \cite{Hoeke2014, Dreiser2013, Mironov2003a} and $[Mn^{II}Mo^{III}]$ complex has $J_z=-34cm^{-1}$ and $J_x=J_y=-11cm^{-1}$ \cite{Mironov2015, Qian2013}. In this study, we employ a generalized ferromagnetic XYZ model for nearest neighbor spin-spin interactions and on-site anisotropy. Using the full Fock space of the Hamiltonian, we follow the properties such as magnetization, susceptibility and specific heat of spin chains with ferromagnetic interaction and different site spins. In the next section we discuss briefly spin Hamiltonian we have studied and present the numerical approach for obtaining the properties of the model. In the third section, we present the result of a purely anisotropic exchange model. This will be followed by the results on a model with both exchange and site anisotropies in section four. We will end the paper with a discussion of all the results.
  
\section{\label{sec:methodology}Methodology}
The basic starting Hamiltonian for studying most magnetic materials is the isotropic Heisenberg exchange model given by
\begin{eqnarray}
\label{eqn:Heisenberg}
\hat{\mathcal{H}}_{Heis}=\sum_{\langle i,j\rangle} J_{ij} \hat{S}_i\cdot \hat{S}_j
\end{eqnarray}
where the summation is over nearest neighbors. This model assumes that spin-orbit interactions are weak and hence the exchange constant $J$ associated with the three components of the spin are equal ($J_{ij}^x=J_{ij}^y=J_{ij}^z$). The isotropic model conserves both total $M_s$ and total $S$ and hence we can choose a spin adapted basis such as the valence bond (VB) basis to set up the Hamiltonian matrix. The Rumer-Pauling VB basis is nonorthogonal and hence the Hamiltonian matrix is nonsymmetric. While computing eigenstates of a nonsymmetric matrix is reasonably straight forward, computing properties of the eigenstates in the VB basis is nontrivial. However, the VB eigenstates can be transformed to eigenstates in constant $M_s$ basis and the latter basis being orthonormal is easily amenable to computing properties of the eigenstates.

When the spin-orbit interactions are weak, we can include the anisotropy arising from it by adding the site anisotropy term,
\begin{eqnarray}
\label{eqn:AnisHam}
\hat{\mathcal{H}}_{aniso}=\sum_{i} [d_{i,z} \hat{s}_{i,z}^2+d_{i,x} \hat{s}_{i,x}^2+d_{i,y} \hat{s}_{i,y}^2]
\end{eqnarray}
($d_{i,x}$, $d_{i,y}$ and $d_{i,z}$ are local ion anisotropies) and treating it as a perturbation. Usually, it is sufficient to deal with just the site diagonal anisotropy and set  $d_{i,x}=d_{i,y}=0$. However, if the local anisotropy axis is not aligned with the global spin axis, then we need to include the off-diagonal site anisotropy terms. For weak on-site anisotropy ($\frac{d}{J} << 1$), we can obtain the splitting of a given total spin state perturbatively by determining the molecular anisotropy parameters $D_M$ and $E_M$ given by the eigenstates of the Hamiltonian in a given spin state $S$ \cite{Raghunathan2008},
\begin{eqnarray}
\label{eqn:Molanis}
\hat{\mathcal{H}}_{mol}&=&D_M \left( \hat{S}_z^2 - \frac{1}{3}S(S+1) \right)+E_M(\hat{S}_x^2-\hat{S}_y^2)
\end{eqnarray}
Spin-orbit interaction can also lead to anisotropy in the exchange Hamiltonian leading to a general $XYZ$ model whose Hamiltonian is given by 
\begin{eqnarray}
\label{eqn:XYZHaml}
\hat{\mathcal{H}}_{XYZ}=\sum_{\langle ij \rangle} [J_{ij}^x \hat{s}_i^x\cdot \hat{s}_j^x+J_{ij}^y \hat{s}_i^y\cdot \hat{s}_j^y+J_{ij}^z \hat{s}_i^z\cdot \hat{s}_j^z]
\end{eqnarray} 
for $J_{ij}^x \neq J_{ij}^y \neq J_{ij}^z$. In this model, there does not exist any spin symmetry and we need to solve the Hamiltonian for its eigenstates in the full Fock basis with no restrictions on total $S$ or $M_s$. In cases where a system has the same exchange constant along x and y directions but different from the exchange constant in the z-direction, we obtain the XXZ model with the Hamiltonian is given by 
\begin{eqnarray}
\label{eqn:XXZHaml}
\hat{\mathcal{H}}_{XXZ}=\sum_{\langle ij \rangle} J_{ij}^x [\hat{s}_i^x\cdot \hat{s}_j^x+\hat{s}_i^y\cdot \hat{s}_j^y]+J_{ij}^z \hat{s}_i^z\cdot \hat{s}_j^z
\end{eqnarray} 

\noindent For convenience we write the general XYZ Hamiltonian in eqn. \ref{eqn:XYZHaml} as 
\begin{eqnarray}
\label{eqn:generalHaml}
\hat{\mathcal{H}}=\sum_{\langle ij \rangle} J [\hat{s}_i^z\cdot \hat{s}_j^z+(\gamma + \delta)\hat{s}_i^x\cdot \hat{s}_j^x+(\gamma - \delta) \hat{s}_i^y\cdot \hat{s}_j^y]
\end{eqnarray}

\noindent where $J_{ij}^z=J$, $\gamma = \frac{J_{ij}^x+J_{ij}^y}{2J}$ and $\delta = \frac{J_{ij}^x-J_{ij}^y}{2J}$. The deviation of $\frac{J_{ij}^x+J_{ij}^y}{2}$ from $J_{ij}^z$ is then represented by the parameter $\epsilon = 1-\gamma$ and the difference between exchange along x and y directions in normalized units is $\delta$. This model can be solved in the constant $M_s$ basis. Besides exchange anisotropy, a system can also have site anisotropy in which case, the $\hat{\mathcal{H}}_{aniso}$ should be considered together with the respective Hamiltonian, either perturbatively (for weak on site anisotropy) or in the zeroth order Hamiltonian itself.

The effect of large anisotropic exchange or large site anisotropy is to mix states with different total spin $S$. Thus, the conventional approach to define molecular anisotropy constants through the effective Hamiltonian (eqn. \ref{eqn:Molanis}) fails as the low-lying multiplet states can not be identified as arising from a unique total spin state, as, the total spin of a state is not conserved. In such situations, the approach we have taken is to obtain the thermodynamic properties such as susceptibility $\chi(T)$, magnetization $M(T)$ and specific heat $C_v(T)$ of the system as a function of Hamiltonian parameters. These are computed from the canonical partition function obtained from the full spectrum of the Hamiltonian. The full Fock space of the Hamiltonian is given by $(2s_i+1)^N$, where N is the number of sites in the spin chain. The largest system we have studied corresponds to $s_i=2$ and $N=5$ which spans a Fock space of dimensionality of 3,125. We need to calculate $\langle \langle M_s \rangle \rangle$ for the magnetic properties which is a thermodynamic average of the expectation values in the eigenstates. To obtain the spin expectation value $\langle \hat S^2 \rangle$ in an eigenstate we have computed the spin-spin correlation functions $\langle \hat s_i^z \hat s_j^z \rangle$, $\langle \hat s_i^x \hat s_j^x \rangle$ and $\langle \hat s_i^y \hat s_j^y \rangle$. 

\section{\label{sec:RandD}Anisotropic Exchange Models}
Here, we discuss the magnetic anisotropy arising only from the exchange anisotropy. In the small exchange anisotropy limit, we first consider the XXZ model and XYZ model with small $\delta$. We will end this section with a discussion of the XYZ models with large anisotropy parameters $\epsilon$ and $\delta$. All the exchange interactions are taken to be ferromagnetic.

\subsection{\label{sec:XXZ}Small Anisotropy models} 
\begin{figure*}
    \includegraphics[width=10cm]{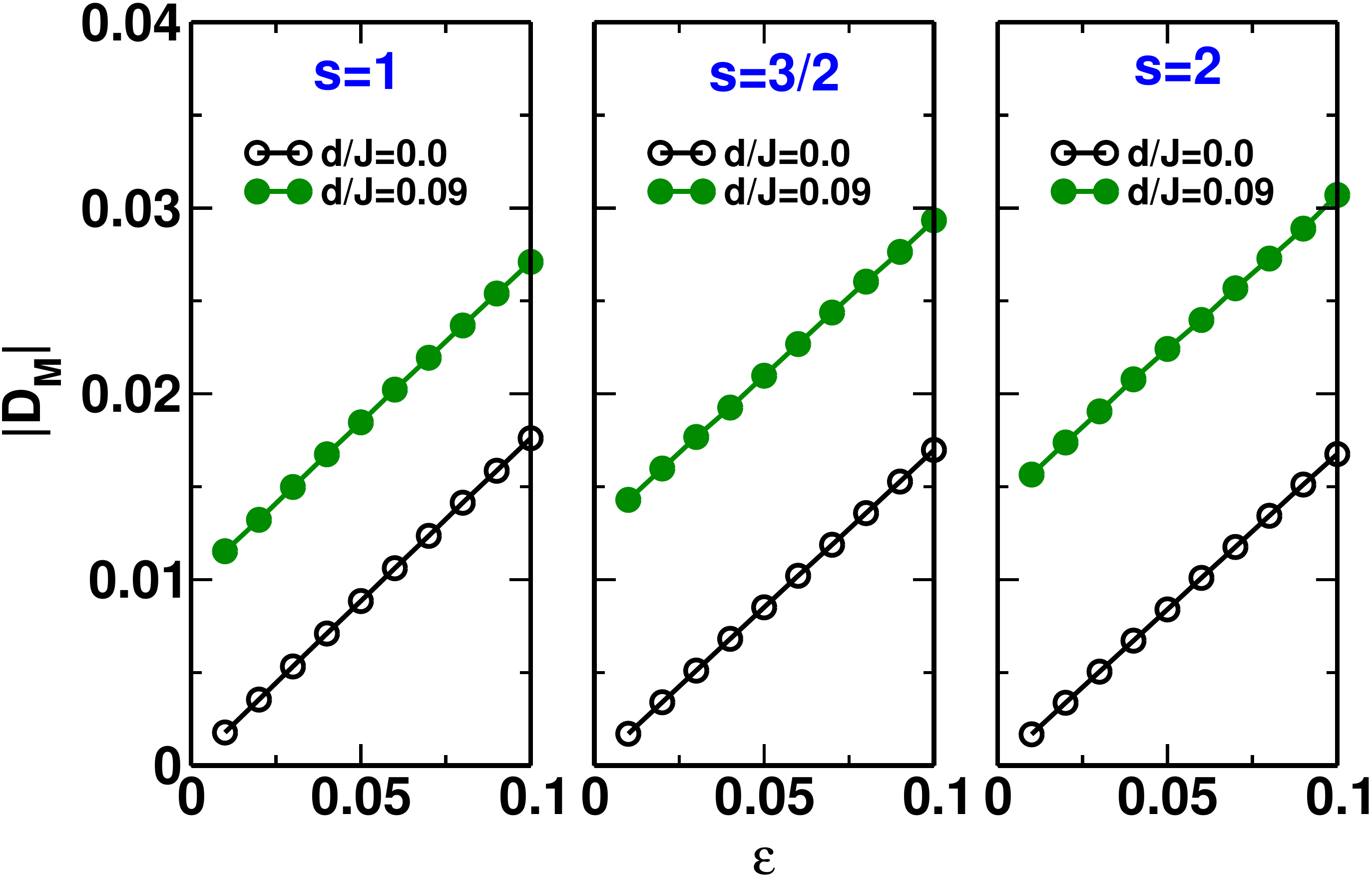} 
      \caption{\label{fig:DMvsepsln}Dependence of molecular anisotropy parameter $|D_M|$ on $\epsilon$ for spin chains with site spins $s=1$, $3/2$ and $2$ and chain length $N=5$. $D_M$ values are computed by fitting the energy gaps to the Hamiltonian $D_M S_{z}^2$ for on-site anisotropy $d/J=0$ and $0.09$. }
\end{figure*}

In this model we set $\delta$ to zero in eqn. \ref{eqn:generalHaml} and study spin chains with site spins $1$, $3/2$ and $2$ in chains of 4 and 5 sites with open boundary condition. We have not considered spin-1/2 system since we wish to study the synergistic effect of anisotropic exchange and on-site anisotropy. The latter exists only for site spin greater than half. The ground state in each case corresponds to $\pm M_s=Ns$ where $N$ is the number of sites and $s$ is the site spin. The total spin of the states is calculated from the eigenstates as expectation value of $\hat S^2$.
\begin{center}
\begin{table}[h!]
\caption{\label{tab:XXZTableFive}Energy gaps (in units of $J$) from the ground state of the low-lying states lying below the lowest state with $M_s$=0. $M_s$ is conserved and is a good quantum number. The total spin $S_{tot}$ is calculated from the expectation value $\langle \hat S^2 \rangle$ of the state. Intruder states are shown in red.}
~~\\
\setlength{\tabcolsep}{1.0pt}
\begin{tabular}{|c|c|c|c|c|c|c|c|c|c|c|c|} \hline  
  & \multicolumn{11}{|c|} {N=5, XXZ model}   \\
 \cline{2-12}
 $\epsilon$ & \multicolumn{3}{|c|} {s=1} & &
 \multicolumn{3}{|c|} {s=3/2} & & 
 \multicolumn{3}{|c|} {s=2} \\
\cline{2-4} \cline{6-8} \cline{10-12}
 &$M_s$ & $S_{tot}$ & Energy && $M_s$ & $S_{tot}$ & Energy && $M_s$ & $S_{tot}$ & Energy\\ \hline
~ 0.10 ~&~ $\pm 5$ ~&~ 5.00 ~&~ 0 ~&&~ $\pm 7.5$ ~&~ 7.50 ~&~ 0 ~&&~ $\pm 10$ ~&~ 10.00 ~&~ 0 ~\\
        & $\pm 4$ & 4.99 &~0.158~&& $\pm 6.5$ & 7.49 &~0.237~&& $\pm 9$ &  9.99 &~0.316~\\
        & $\pm 3$ & 4.99 &~0.282~&& $\pm 5.5$ & 7.49 &~0.442~&& $\pm 8$ &  9.99 &~0.601~\\
        & $\pm 2$ & 4.99 &~0.370~&& $\pm 4.5$ & 7.49 &~0.612~&& $\pm 7$ &  9.99 &~0.852~\\
        & $\pm 1$ & 4.99 &~0.423~&&\textcolor{red}{$\pm 6.5$}&\textcolor{red}{6.50}&\textcolor{red}{0.706}&&\textcolor{red}{$\pm 9$}&\textcolor{red}{9.00}&\textcolor{red}{0.941}\\
        &   0  & 4.99 &~0.441~&& $\pm 3.5$ & 7.49 &~0.749~&& $\pm 6$ &  9.99 &~1.071~\\
     &      &       &         && $\pm 2.5$ & 7.49 &~0.852~&&\textcolor{red}{$\pm 8$}&\textcolor{red}{8.99}&\textcolor{red}{1.217}\\
     &      &       &         &&\textcolor{red}{$\pm 5.5$}&\textcolor{red}{6.49}&\textcolor{red}{0.902}&& $\pm 5$ &  9.99 &~1.256~\\
     &      &       &         && $\pm 1.5$ & 7.49 &~0.921~&& $\pm 4$ &  9.99 &~1.407~\\ 
     &      &       &         && $\pm 0.5$ & 7.49 &~0.955~&&\textcolor{red}{$\pm 6$}&\textcolor{red}{8.99}&\textcolor{red}{1.462}\\
     &      &       &         &&          &       &         &&\textcolor{red}{$\pm 7$}&\textcolor{red}{9.99}&\textcolor{red}{1.526}\\
     &      &       &         &&          &       &         && $\pm 3$ &  9.99 &~1.611~\\
     &      &       &         &&          &       &         && $\pm 2$ &  9.99 &~1.661~\\
     &      &       &         &&          &       &         && $\pm 1$ &  9.99 &~1.673~ \\ 
     &      &       &         &&          &       &         &&   0  &  9.99 &~1.678~ \\ \hline 
\hline     
0.15 ~&~ $\pm 5$ ~&~ 5.00 ~&~ 0 ~&&~ $\pm 7.5$ ~&~ 7.50 &~ 0 ~&&~ $\pm 10$ ~&~ 10.00 ~&~ 0 ~\\
      &  $\pm 4$  &  4.99  &~ 0.236 ~&&  $\pm 6.5$  &  7.49 &~ 0.354~&&   $\pm 9$  &   9.99  &~ 0.472 ~\\
      &  $\pm 3$  &  4.99  &  0.420  &&  $\pm 5.5$  &  7.49 &  0.658  &&   $\pm 8$  &   9.99  &  0.895 \\
      &\textcolor{red}{$\pm 4$}&\textcolor{red}{4.00}&\textcolor{red}{0.514}&&\textcolor{red}{$\pm 6.5$}&\textcolor{red}{6.50}&\textcolor{red}{0.771}&&\textcolor{red}{$\pm 9$}&\textcolor{red}{8.99}&\textcolor{red}{1.028} \\
      &  $\pm 2$  &  4.98  &  0.551  &&  $\pm 4.5$  &  7.489 &  0.914  &&   $\pm 7$   &   9.991  & 1.272  \\
      &  $\pm 1$  &  4.98  &  0.630  &&\textcolor{red}{$\pm 5.5$}&\textcolor{red}{6.49}&\textcolor{red}{1.063}&&\textcolor{red}{$\pm 8$}&\textcolor{red}{8.99}&\textcolor{red}{1.441} \\
      &     0   &  4.97  &  0.656  &&  $\pm 3.5$  &  7.48 & 1.118  &&   $\pm 6$   &   9.98  &  1.598 \\
      &         &         &           &&  $\pm 2.5$  &  7.48 &  1.273  &&\textcolor{red}{$\pm 7$}&\textcolor{red}{8.99}&\textcolor{red}{1.805} \\
      &         &         &           &&\textcolor{red}{$\pm 4.5$}&\textcolor{red}{6.49}&\textcolor{red}{1.305}&&  $\pm 5$   &   9.98  & 1.876 \\
      &         &         &           &&  $\pm 1.5$  &  7.48 &  1.375  &&\textcolor{red}{$\pm 8$}&\textcolor{red}{8.00}&\textcolor{red}{2.050} \\
      &         &         &           &&  $\pm 0.5$  &  7.48 &  1.426  &&   $\pm 4$   &   9.98  &  2.104 \\
      &         &         &           &&             &        &           &&\textcolor{red}{$\pm 6$}&\textcolor{red}{8.98}&\textcolor{red}{2.120} \\
      &         &         &           &&             &        &           &&   $\pm 3$   &   9.98  &  2.282 \\
      &         &         &           &&             &        &           &&\textcolor{red}{$\pm 5$}&\textcolor{red}{8.97}&\textcolor{red}{2.386} \\
      &         &         &           &&             &        &           &&\textcolor{red}{$\pm 7$}&\textcolor{red}{8.00}&\textcolor{red}{2.399} \\
      &         &         &           &&             &        &           &&   $\pm 2$   &   9.98  &  2.408 \\
      &         &         &           &&             &        &           &&   $\pm 1$  &   9.98  &  2.484 \\
      &         &         &           &&             &        &           &&      0    &   9.98  &  2.510 \\ \hline
\end{tabular}
\end{table}
\end{center}
In table \ref{tab:XXZTableFive} we present the energy gaps from the ground state of the low-lying states up to first $M_s=0$ state of short spin chains of length up to five spins for different $\epsilon$ values. The table for spin chains of four spins is given in supporting material. We notice from the table that for $\epsilon=0.1$, the lowest energy states of the $s=1$ chains satisfies $E(|M_s|=Ns) < E(|M_s|=Ns-1)...< E(|M_s|=0)$ and the total spin of these states is also very close to $Ns$. In this case we can fit the energy gaps to the Hamiltonian $D_M S_{z}^2$. The diagonal anisotropy of these states is shown in fig. \ref{fig:DMvsepsln}.
In the XXZ model we do not have off-diagonal anisotropy, i.e., $E_M=0$ in the anisotropic Hamiltonian given by eqn. \ref{eqn:Molanis}. We note in table \ref{tab:XXZTableFive} that for spin chains with $s=3/2$ and $s=2$, there are intruder states within the manifold of $S \simeq 7.5$ and $\simeq 10$ respectively. We also find that as $\epsilon$ is increased to $0.15$, even the $s=1$ spin chain has intruders. Furthermore, for site spin $2$, the intruders within the $S=10$ manifold are from progressively lower total spin states, namely $S=9$, $8$ and $7$. Thus, it is not meaningful anymore to define molecular $D_M$ and $E_M$ parameters. For the $N=4$ chains the intruder states occur in $s=1$ chain for $\epsilon=0.25$ and for $s=3/2$ and $s=2$ chain for $\epsilon=0.20$ (see supporting material). Thus, intruders arise at smaller $\epsilon$ values for longer chains and higher site spin. The $|D_M|$ increases linearly with increase in anisotropy (fig. \ref{fig:DMvsepsln}). 
\begin{figure*}
    \includegraphics[width=10cm]{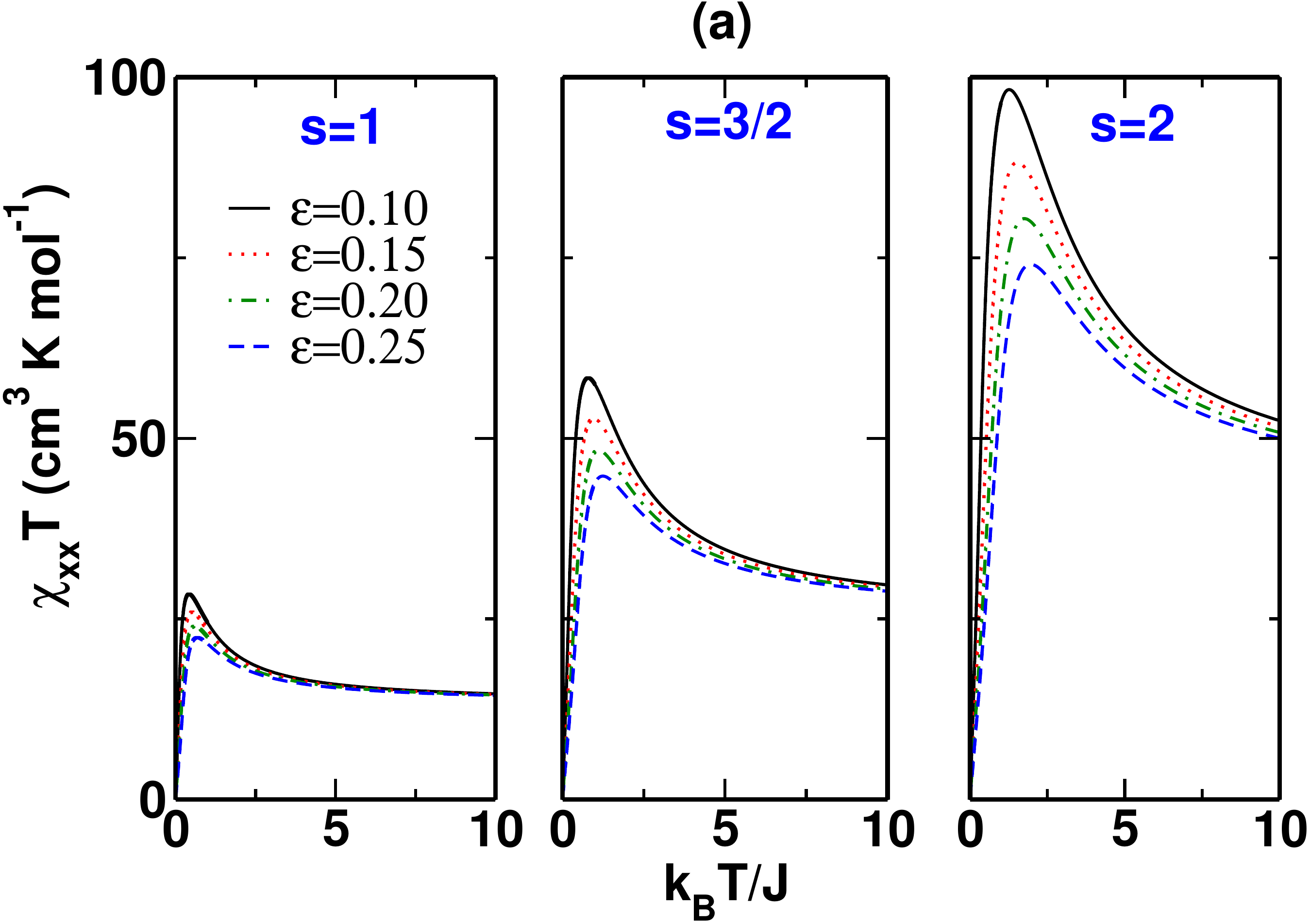}  
    
    \vspace{0.3in}
\includegraphics[width=10cm]{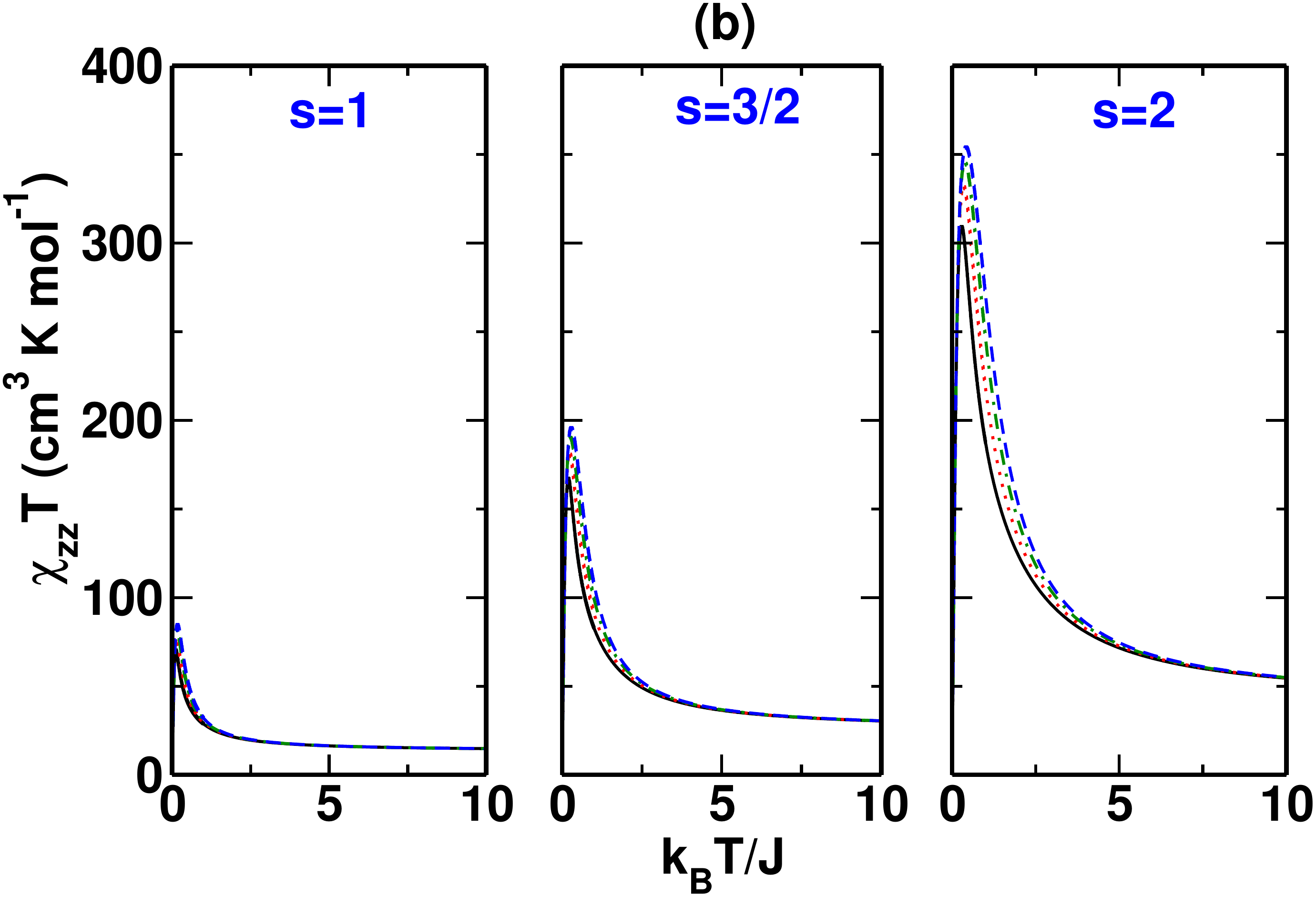} 
      \caption{\label{fig:ChiXX,ChiZZ}(a) Plot of $\chi_{_{xx}}T$ as a function of temperature computed with applied field along x-direction, (b) Plot of $\chi_{_{zz}}T$ as a function of temperature computed with applied field along z-direction for different values of exchange anisotropy $\epsilon$, in the absence of on-site anisotropy. The susceptibilities are computed for field magnitude $H=J/g\beta=0.005$. Color coding and line type is the same for all panels. (Note the scale on y-axis are different for (a) and (b)).}
\end{figure*}

We have obtained the thermodynamic properties of these spin chains as a function of temperature and the magnetization as a function of magnetic field at a fixed temperature. We show in fig. \ref{fig:ChiXX,ChiZZ}, $\chi_{_{xx}}T(=\chi_{_{yy}}T)$ and $\chi_{_{zz}}T$ dependence on temperature for spin chains of five spins for different values of the site spins. Expectedly the susceptibility increases with site spin in all cases. The $\chi_{_{zz}}T$ component is much larger than the $\chi_{_{xx}}T$ component and both show a maxima. The maxima is at a higher temperature for $\chi_{_{xx}}T$ compared to $\chi_{_{zz}}T$ and the $\chi_{_{xx}}T$ maxima is also broader. We also note that $\chi_{_{zz}}T$ is larger than $\chi_{_{xx}}T$ by a factor of between 2 and 3, even though maximum anisotropy $\epsilon$ is only 0.25. Besides the temperature of the maxima also increases with site spin. The $ZZ$ component is larger for large anisotropy while the $XX$ component is smaller at large anisotropy. This is because as $\epsilon$ increases it becomes easier to magnetize along the z-axis, while it becomes harder to magnetize in the x-y plane. This trend is also seen in the magnetization plots as a function of the magnetic field shown in fig. \ref{fig:MxHx,MzHz}. We note that the magnetization $\langle M_z \rangle$ increases with $\epsilon$ while $\langle M_x \rangle$ decreases with $\epsilon$ for the same applied field.
\begin{figure*}
    \includegraphics[width=10cm]{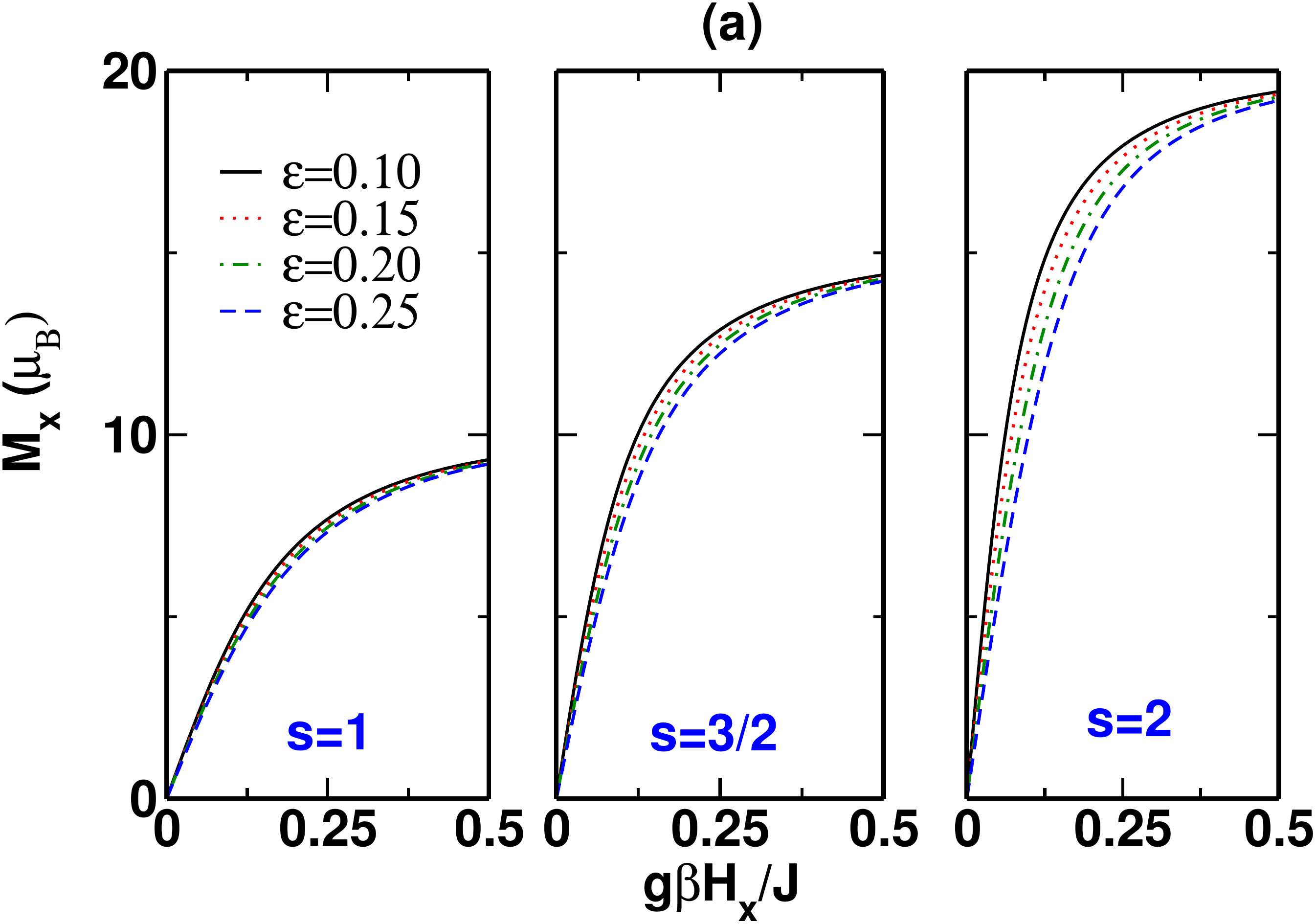}  

\vspace{0.3in}
\includegraphics[width=10cm]{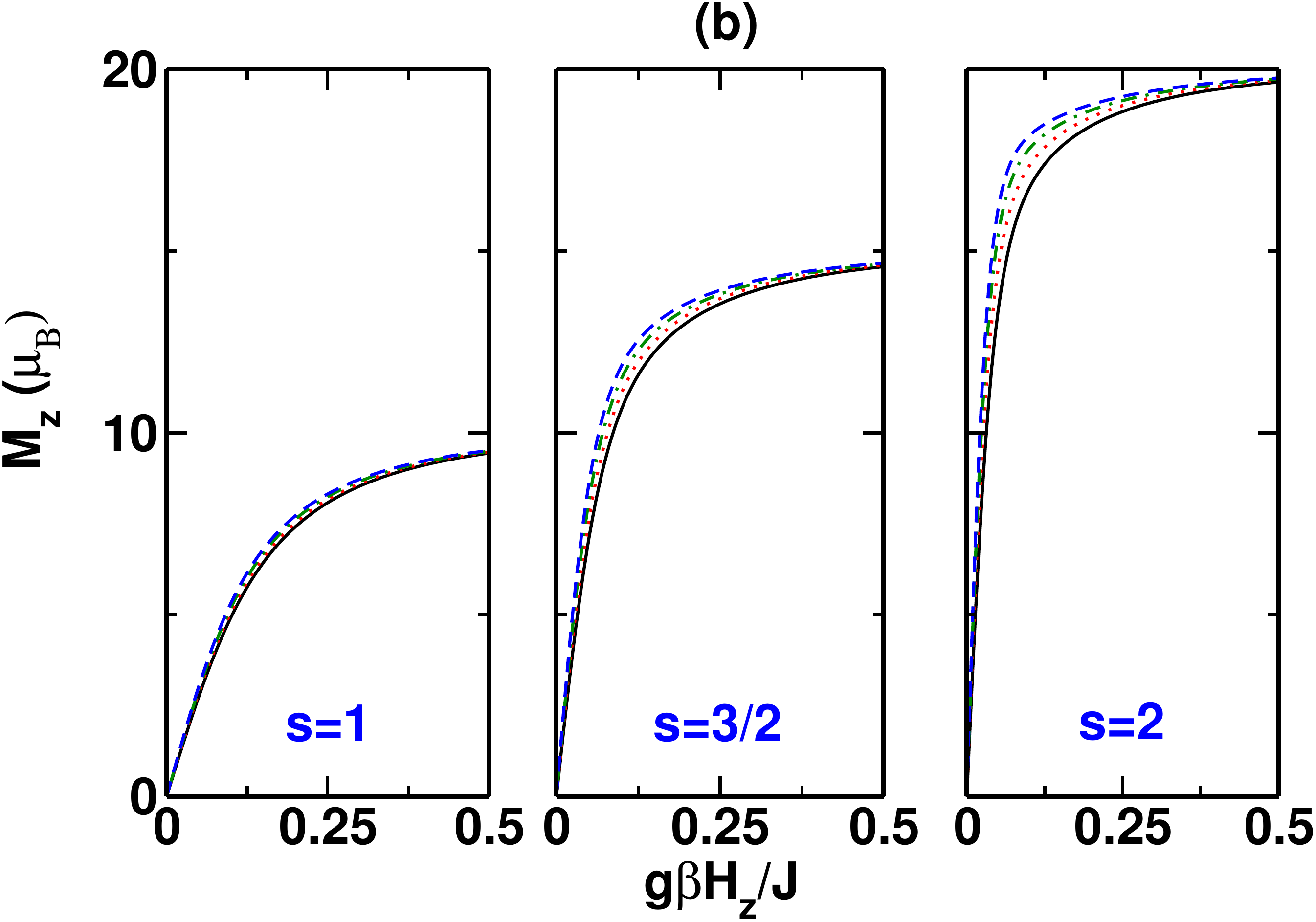} 
      \caption{\label{fig:MxHx,MzHz}Dependence of magnetization ($M$) on applied magnetic field ($g\beta H/J$) at temperature $k_BT/J=1.0$. (a) $M_x$ versus $g\beta H_x/J$, (b) $M_z$ versus $g\beta H_z/J$ for different values of exchange anisotropy $\epsilon$, in the absence of on-site anisotropy.  Color coding and line type is the same for all panels. }
\end{figure*}

\begin{figure*}
    \includegraphics[width=12cm]{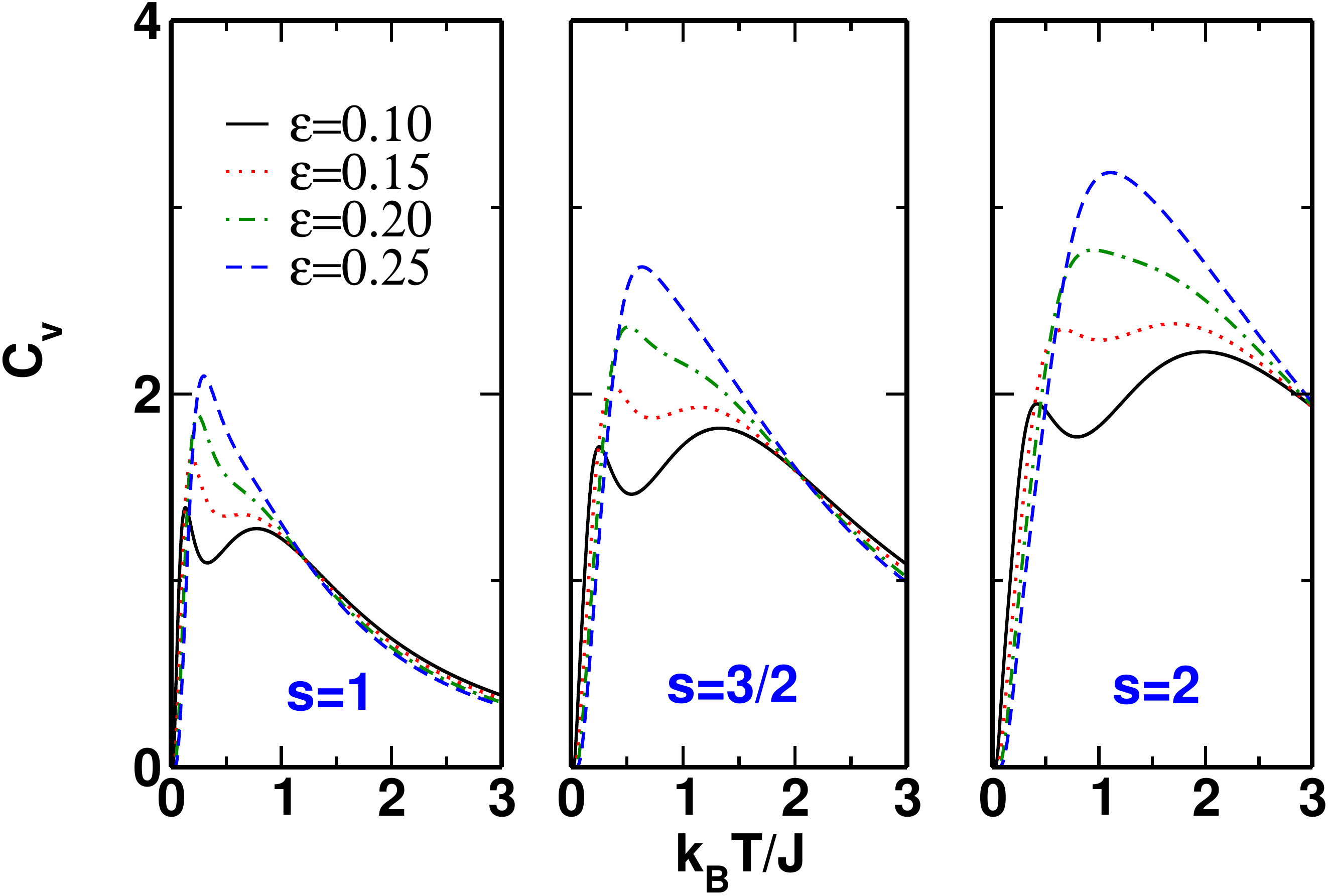} 
      \caption{\label{fig:Cvxxz}Dependence of specific heat ($C_v$) on temperature ($k_BT/J$) of spin chains with $s=1$, $3/2$ and $s=2$ and systems size $N=5$ for different values of axial exchange anisotropy $\epsilon$, in the absence of on-site anisotropy. Color coding and line type is the same for all panels.}
\end{figure*}

\begin{figure*}
\center
    \includegraphics[height=10cm, width=12cm]{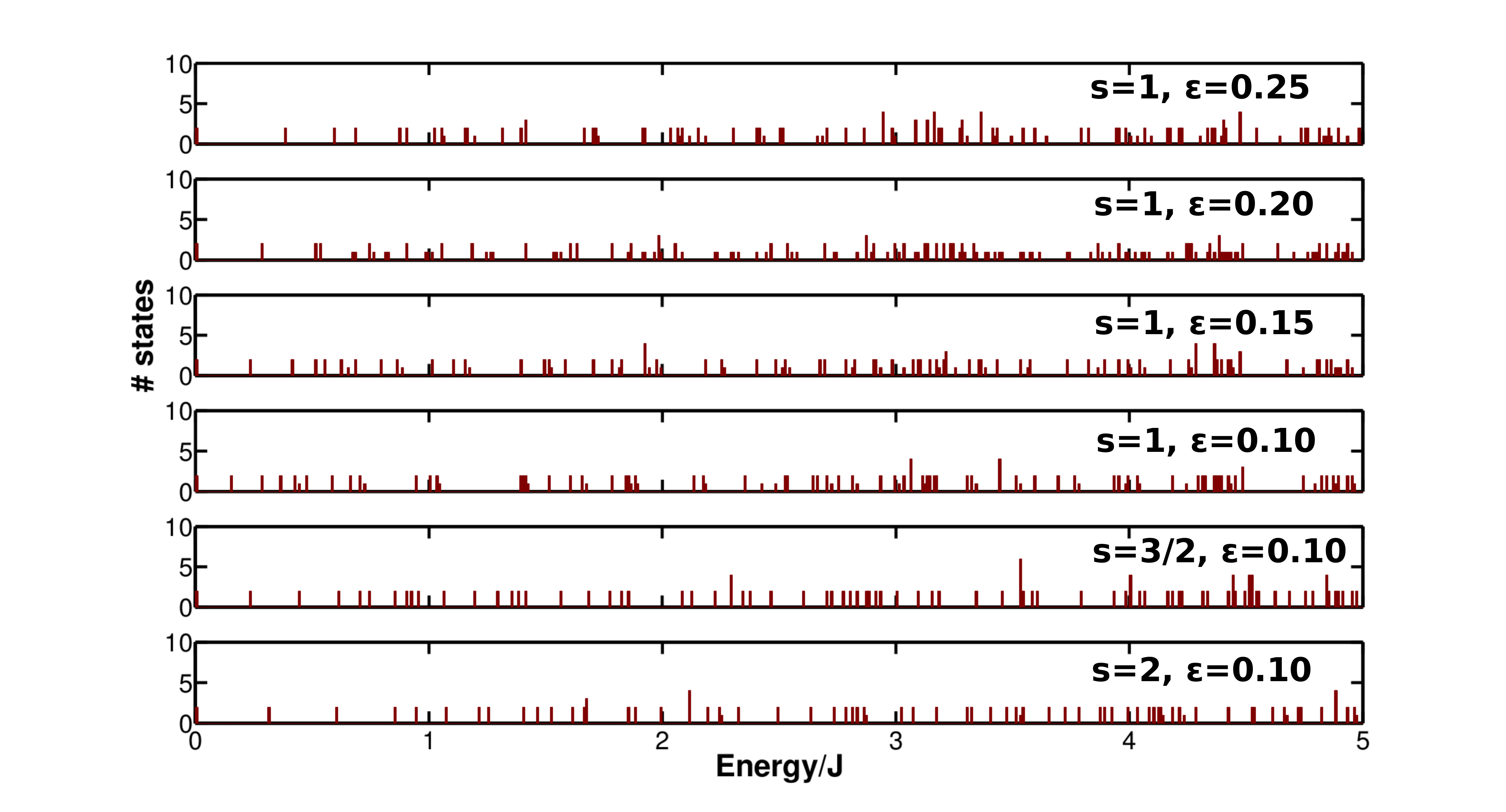} 
      \caption{\label{fig:Histogram}Histogram of number of low-energy states versus energy of all the eigenstates of spin chains with $s=1$, $3/2$ and $2$ and systems size $N=5$. Values of axial anisotropy $\epsilon$ and site spin are shown in each box.}
\end{figure*}

The dependence of specific heat, $C_v$, on temperature for different $\epsilon$ values is shown in fig.  \ref{fig:Cvxxz}.
We find that for small $\epsilon$, the specific heat shows two peaks, the first peak is narrow and the second peak is broad. This is seen for all site spins. This can be understood from the nature of the full energy spectrum of the Hamiltonian for different $\epsilon$ values fig.  \ref{fig:Histogram}. We see that there are two successively small gaps in the spectrum below 0.13J for small anisotropy but these gaps shift to much higher energies for large anisotropy. This implies that at small anisotropy, the specific heat first increases with increase in temperature and then drops as thermal energy can not access higher energy states. As the temperature increases further the higher energy states are populated leading to increase in specific heat. Thus, the magnetic specific heat dependence on temperature can be used as a tool to estimate the anisotropy of the chain.

Introducing small planar anisotropy, $\delta$, does not significantly change the low energy spectrum in table \ref{tab:xyztable6} and consequently there is no discernible change in the thermodynamic properties. The main difference is that $M_s$ is also not conserved even for small values of $\delta$.

\begin{table}
\caption{\label{tab:xyztable6}Energy gaps from the ground state (in units of $J$) of the low-lying states lying within the manifold of spin $S\simeq ns$. Both $M_s$ and $S$ are not conserved and not good quantum numbers. The total spin $S_{tot}$ is calculated from the expectation value $\langle \hat S^2 \rangle$ of the state. Intruder states are shown in red. $\langle M_s \rangle$ are given for states for which it could be computed. $\langle M_s \rangle$ values are not quoted for the states which show large mixing of different $M_s$ states.}
~~\\
\centering
\setlength{\tabcolsep}{1.0pt}
\begin{tabular}{|c|c|c|c|c|c|c|c|c|c|c|c|} \hline  
  & \multicolumn{11}{|c|} {N=5, XYZ model}   \\
 \cline{2-12}
 $\epsilon,\delta$ & \multicolumn{3}{|c|} {s=1} & &
 \multicolumn{3}{|c|} {s=3/2} & & 
 \multicolumn{3}{|c|} {s=2} \\
\cline{2-4} \cline{6-8} \cline{10-12}
 &$M_s$ & $S_{tot}$ & Energy && $M_s$ & $S_{tot}$ & Energy && $M_s$ & $S_{tot}$ & Energy\\ \hline
~0.095,~&~ -4.99, 4.99 ~&~ 4.99 ~&~ 0 ~&&~ -7.49, 7.49~ &~ 7.49 ~&~ 0 ~&&~ -9.99, 9.99 ~&~ 9.99 ~&~ 0 ~\\
0.005      &~-3.99, 3.99~&~4.998~&~0.150~&&~-6.49, 6.49~&~7.49~&~ 0.225~&&~-8.997, 8.99~&~9.99~&~0.301~\\
           &~-2.99, 2.99~&~4.997~&~0.267~&&~-5.49, 5.49~&~7.49~&~0.419~&&~-7.995, 7.99~&~9.99~&~0.570~\\
           &~-1.94, 2.01~&~4.995~&~0.347~&&~-4.48, 4.48~&~7.49~&~0.581~&&~-6.992, 6.99~&~9.99~&~0.808~\\
           &~-1.04, 1.03~&~4.993~&~0.389~&&\textcolor{red}{-6.498, 6.49}&\textcolor{red}{6.49}&\textcolor{red}{0.699}&&\textcolor{red}{-8.99, 8.99}&\textcolor{red}{8.99}&\textcolor{red}{0.932}\\
           &~ \textbf{\textemdash}  ~&~4.99~&~0.423~&&~-3.46, 3.46~&~7.49~&~0.709~&&~-5.98, 5.98~&~9.99~&~1.015~\\
           &~             &~     ~&~        ~&&~-2.33, 2.33~&~7.49~&~0.805~&&~-4.97, 4.97~&~9.99~&~1.189~\\
           &~             &~     ~&~        ~&&~-1.51, 1.51~&~7.49~&~0.865~&&\textcolor{red}{-7.99, 7.99}&\textcolor{red}{8.99}&\textcolor{red}{1.194}\\
           &              &       &         &&\textcolor{red}{-5.49, 5.49}&\textcolor{red}{6.49}&\textcolor{red}{0.885}&&~-3.95, 3.96~&~9.99~&~1.332~\\  
           &              &       &         &&~-0.97, 0.97~&~7.49~&~0.928~&&\textcolor{red}{-6.99, 6.99}&\textcolor{red}{8.99}&\textcolor{red}{1.426}\\
           &              &       &         &&~            ~&~     ~&~       ~&&~-2.82, 2.94~&~9.99~&~1.435~\\ 
           &              &       &         &&~            ~&~     ~&~       ~&&~ \textbf{\textemdash}   ~&~9.99~&~1.497~\\
           &              &       &         &&~            ~&~     ~&~       ~&&~  \textbf{\textemdash}  ~&~9.99~&~1.539~\\
           &              &       &         &&~            ~&~     ~&~       ~&&~  \textbf{\textemdash}  ~&~9.99~&~1.552~\\ 
           &              &       &         &&~            ~&~     ~&~       ~&&\textcolor{red}{-5.98, 5.98}&\textcolor{red}{8.99}&\textcolor{red}{1.626}\\
           &              &       &         &&~            ~&~     ~&~       ~&&~ \textbf{\textemdash}  ~&~9.99~&~1.636~\\
           &              &       &         &&~            ~&~     ~&~       ~&&~ \textbf{\textemdash} ~&~9.99~&~1.637~\\ \hline
\hline
~0.15,     ~&~ -4.95, 4.95 ~&~ 4.99 ~&~ 0 ~&&~ -7.46, 7.46 ~&~ 7.49 ~&~ 0 ~&&~ -9.99, 9.99 ~&~ 9.99 ~&~ 0 ~\\
0.05       &~-3.80, 3.80~&~4.991~&~0.214~&&~-6.37, 6.37~&~7.49~&~0.329~&&~-8.99, 8.99~&~9.99~&~0.441~\\
           &~   -2.57    ~&~4.99~&~0.352~&&~-5.23, 5.23~&~7.49~&~0.605~&&~-7.77, 7.78~&~9.99~&~0.832~\\
           &~    2.57    ~&~4.99~&~0.403~&&\textcolor{red}{-6.46, 6.46}&\textcolor{red}{6.49}&\textcolor{red}{0.770}&&\textcolor{red}{-8.96, 8.96}&\textcolor{red}{8.99}&\textcolor{red}{1.025}\\
           &~    1.74    ~&~4.98~&~0.448~&&~-3.93, 3.93~&~7.48~&~0.817~&&\textcolor{red}{-8.99, 8.99}&\textcolor{red}{8.99}&\textcolor{red}{1.166}\\
           &\textcolor{red}{-3.94, 3.96}&\textcolor{red}{3.99}&\textcolor{red}{0.514}&&~-3.28, 3.28~&~7.48~&~0.992~&&~-6.58, 6.65~&~9.99~&~1.266~\\
           &~    -1.74    ~&~4.98~&~0.588~&&~-3.19, 3.19~&~6.48~&~1.039~&&\textcolor{red}{-7.87, 7.87}&\textcolor{red}{8.99}&\textcolor{red}{1.410}\\
           &~ \textbf{\textemdash} ~&~4.98~&~0.590~&&~-4.15, 4.18~&~7.48~&~1.200~&&~ \textbf{\textemdash}  ~&~9.98~&~1.423~\\
           &\textcolor{red}{\textbf{\textemdash}}&\textcolor{red}{3.98}&\textcolor{red}{0.656}&&\textcolor{red}{-3.03, 3.03}&\textcolor{red}{6.49}&\textcolor{red}{1.251}&&~ \textbf{\textemdash} ~&~9.98~&~1.462~\\  
           &\textcolor{red}{\textbf{\textemdash}}&\textcolor{red}{3.99}&\textcolor{red}{0.675}&&~-2.67, 2.67~&~7.49~&~1.406~&&\textcolor{red}{\textbf{\textemdash}}&\textcolor{red}{8.98}&\textcolor{red}{1.595}\\
           &\textcolor{red}{\textbf{\textemdash}}&\textcolor{red}{3.97}&\textcolor{red}{0.737}&&\textcolor{red}{-5.45, 5.45}&\textcolor{red}{6.49}&\textcolor{red}{1.493}&&~ \textbf{\textemdash} ~&~9.99~&~1.732~\\ 
           & \textbf{\textemdash}  &~4.98~&~0.815~&&~-2.89, 2.89~&~7.48~&~1.529~&&\textcolor{red}{-6.74, 6.76}&\textcolor{red}{8.98}&\textcolor{red}{1.743}\\
           & \textbf{\textemdash}  &~4.98~&~0.816~&&~-4.32, 4.32~&~7.48~&~1.562~&&~ \textbf{\textemdash}  ~&~9.99~&~1.769~\\
           &              &       &         &&\textcolor{red}{-2.48, 2.48}&\textcolor{red}{6.48}&\textcolor{red}{1.744}&&\textcolor{red}{\textbf{\textemdash}}&\textcolor{red}{8.97}&\textcolor{red}{2.013}\\ 
           &              &       &         &&~            ~&~     ~&~       ~&&~ \textbf{\textemdash}  ~&~9.98~&~2.026~\\
           &              &       &         &&~            ~&~     ~&~       ~&&\textcolor{red}{\textbf{\textemdash}}&\textcolor{red}{8.97}&\textcolor{red}{2.027}\\
           &              &       &         &&~            ~&~     ~&~       ~&&~ \textbf{\textemdash}  ~&~9.99~&~2.030~\\ \hline
\end{tabular}
\end{table}

\subsection{\label{sec:XYZ}Large Anisotropy models}
To explore the properties of the spin chains in the large anisotropy limit, we have studied $s=1$, $3/2$ and $2$ models with $\epsilon$ up to $0.75$ and $\delta$ up to $0.15$. In this limit, there are no conserved spin quantities, hence we have studied only thermodynamic properties by computing thermodynamic averages from expectation values in the eigenstates of the Hamiltonian.

All the three diagonal components of the susceptibility as a function of temperature are shown in fig. \ref{fig:ChiXX,chiYY,ChiZZ}. We find that for large anisotropy $\chi_{_{zz}}T$ increases with $\epsilon$ and $\delta$, while $\chi_{_{xx}}T$ and $\chi_{_{yy}}T$ decreases with $\epsilon$ and $\delta$. $\chi_{_{zz}}T$ shows a smooth maxima for all cases we have studied but $\chi_{_{xx}}T$ and $\chi_{_{yy}}T$ do not show a discernible maxima. The $\chi_{_{zz}}T$ maxima occur at lower temperature than  $\chi_{_{xx}}T$ and $\chi_{_{yy}}T$ maxima (when they exist). More significantly $\chi_{_{zz}}T$ is higher for higher anisotropy while $\chi_{_{xx}}T$ and $\chi_{_{yy}}T$ are higher for lower anisotropy.

\begin{figure*}
\center
    \includegraphics[width=8cm]{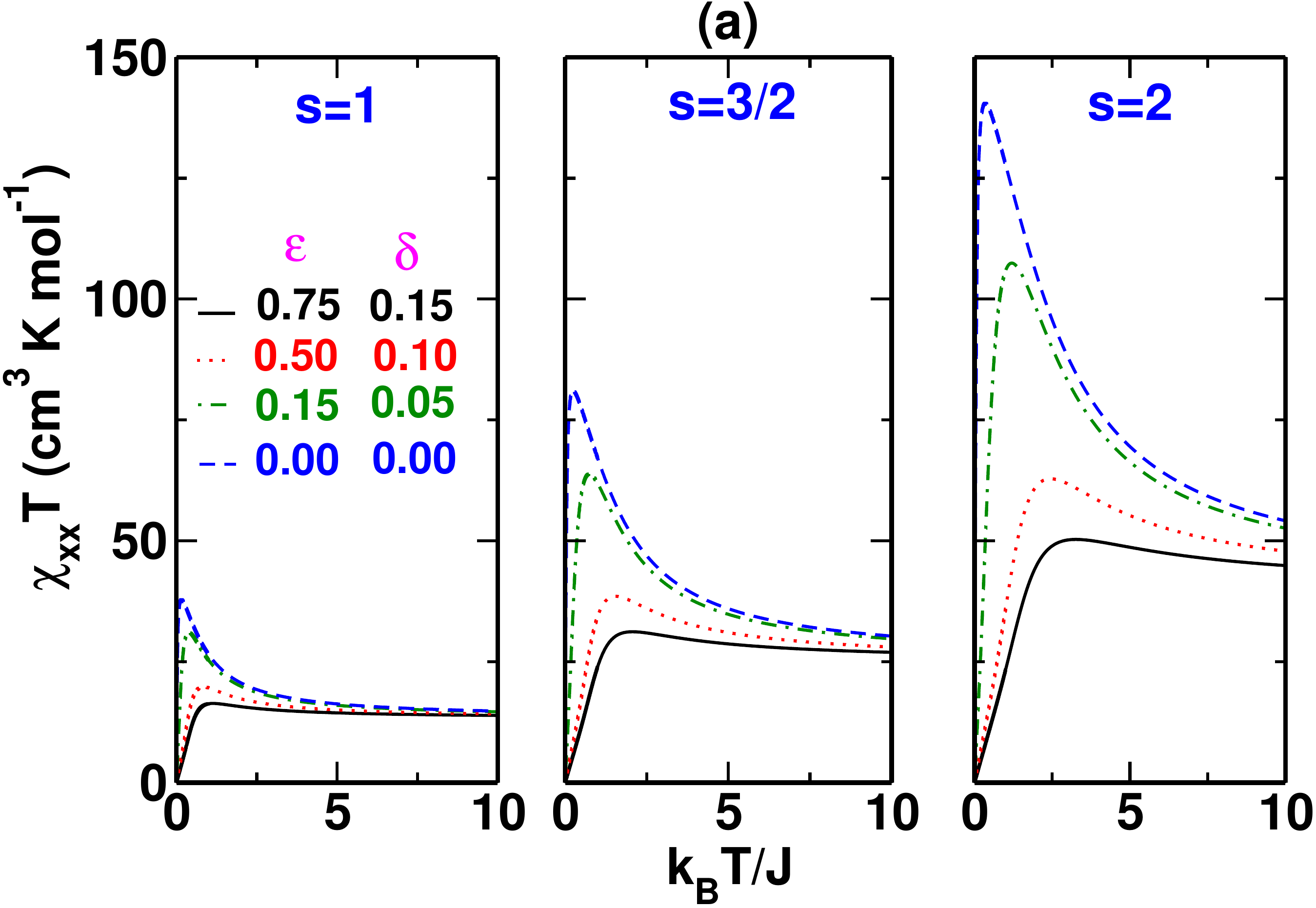}
 \center
     \includegraphics[width=8cm]{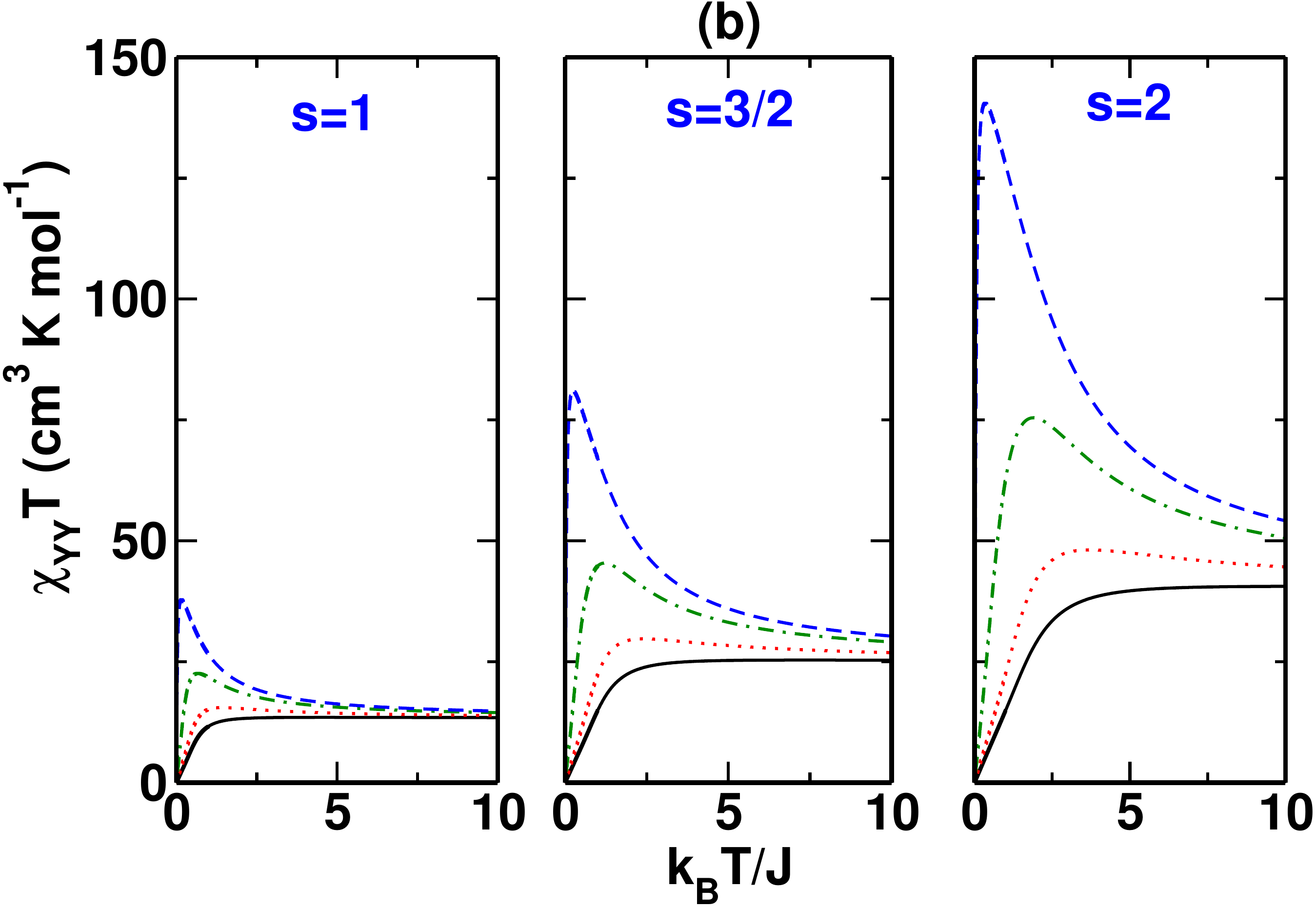}  
\center
     \includegraphics[width=8cm]{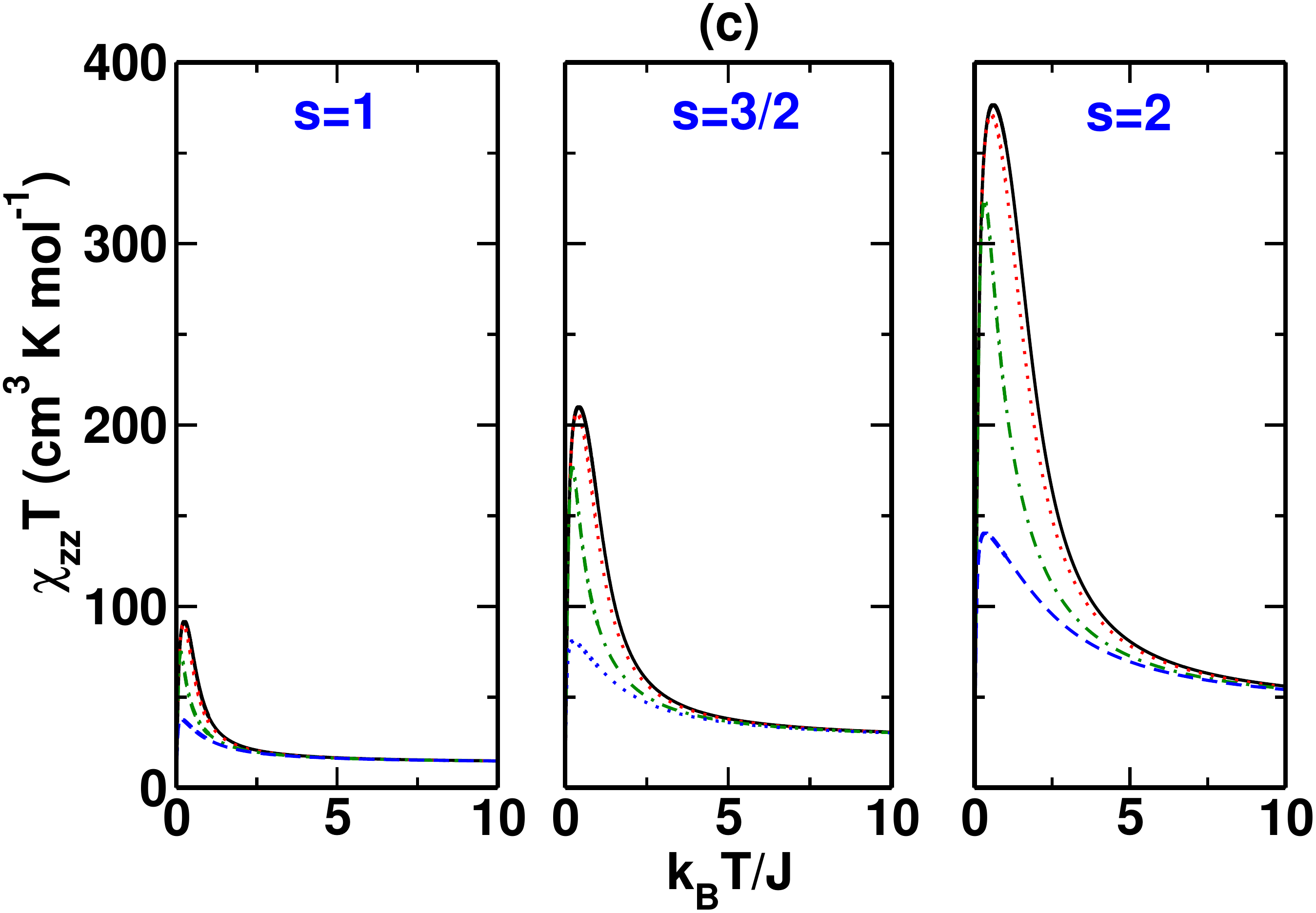} 
      \caption{\label{fig:ChiXX,chiYY,ChiZZ}(a) Plot of $\chi_{_{xx}}T$ as a function of temperature computed with applied field along x-direction ($H_x=J/g\beta=0.005$), (b) Plot of $\chi_{_{yy}}T$ as a function of temperature computed with applied field along y-direction ($H_y=J/g\beta =0.005$) and (c) Plot of $\chi_{_{zz}}T$ as a function of temperature computed with applied field along z-direction ($H_z=J/g\beta=0.005$) for different values of $\epsilon$ and $\delta$, in the absence of on-site anisotropy. Color coding and line type is the same for all panels. (Note scale for (c) are different from those of (a) and (b)). }
\end{figure*}
\begin{figure*}
    \includegraphics[width=12cm]{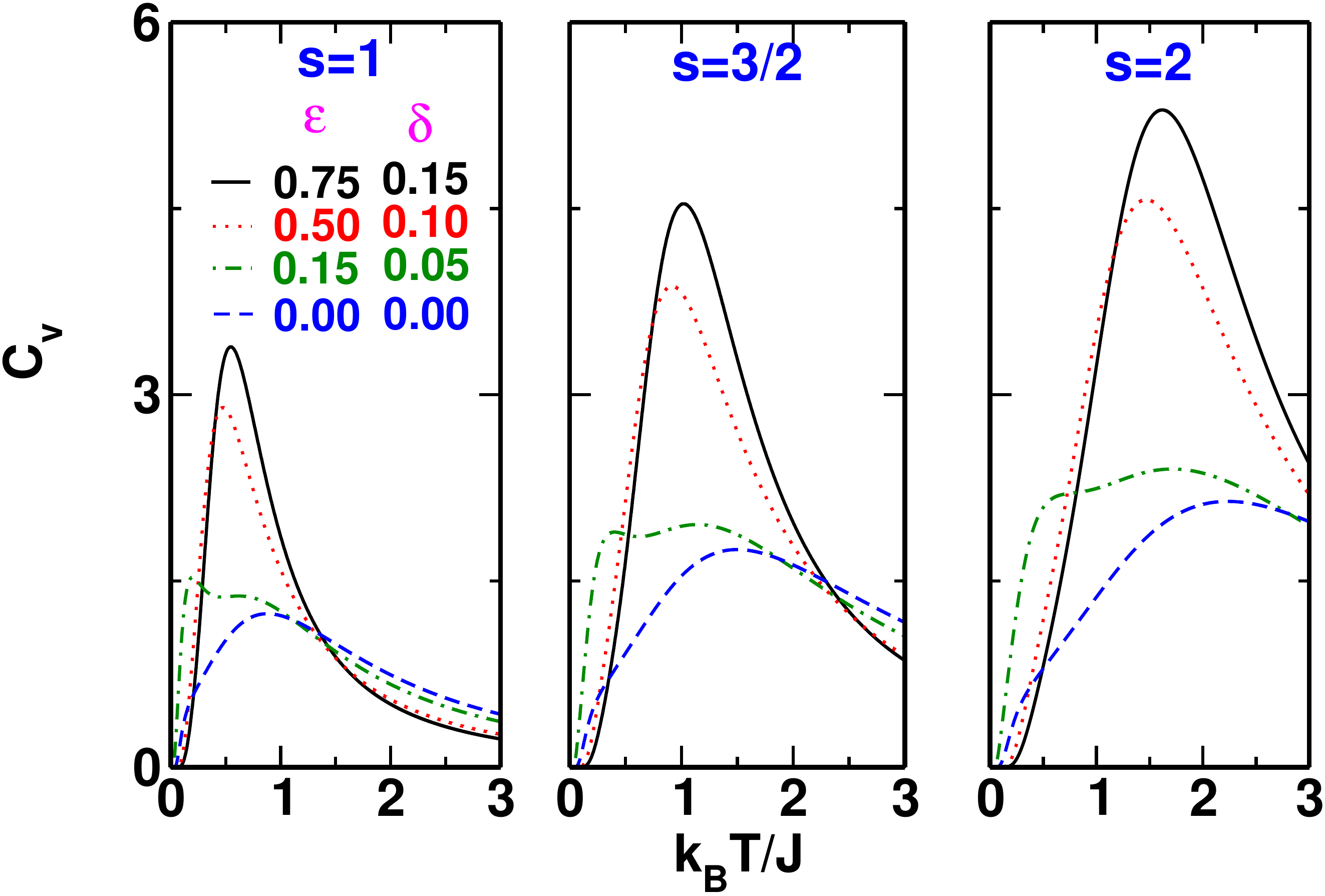} 
      \caption{\label{fig:Cvxyz}Dependence of specific heat ($C_v$) on temperature ($k_BT/J$) of spin chains with $s=1$, $3/2$ and $s=2$ and systems size $N=5$ for different values of $\epsilon$ and $\delta$, in the absence of on-site anisotropy. Color coding for $2^{nd}$ and $3^{rd}$ panel are same as for the first panel.}
\end{figure*}

In fig. \ref{fig:MxHx,MyHy,MzHz} we show the behaviour of magnetization as a function of the field at $k_BT/J=1$. We find very different behaviour for $M_z$ compared to $M_x$ or $M_y$. The $M_z$ component shows saturation at low magnetic fields. The saturation field decreases with increasing site spin. On the other hand, the $M_x$ and $M_y$ components show saturation only for small anisotropy. For large anisotropy they do not show saturation and show a nearly linear increase in magnetization component over the full range of the applied magnetic field. Furthermore, the magnitude of the magnetization decreases with increasing anisotropy at a given field strength.
\begin{figure*}
\center
    \includegraphics[width=8cm]{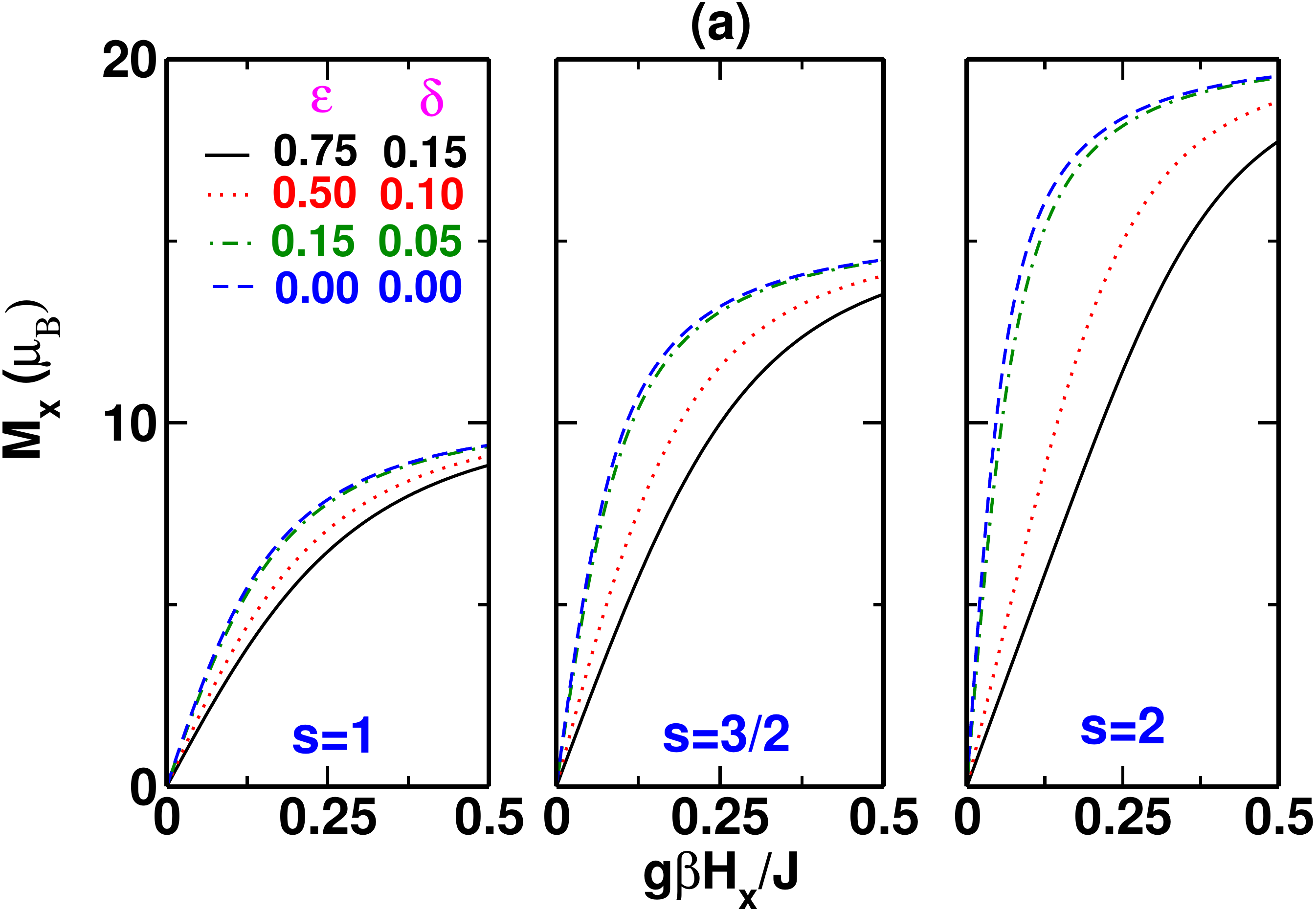} 
\center
    \includegraphics[width=8cm]{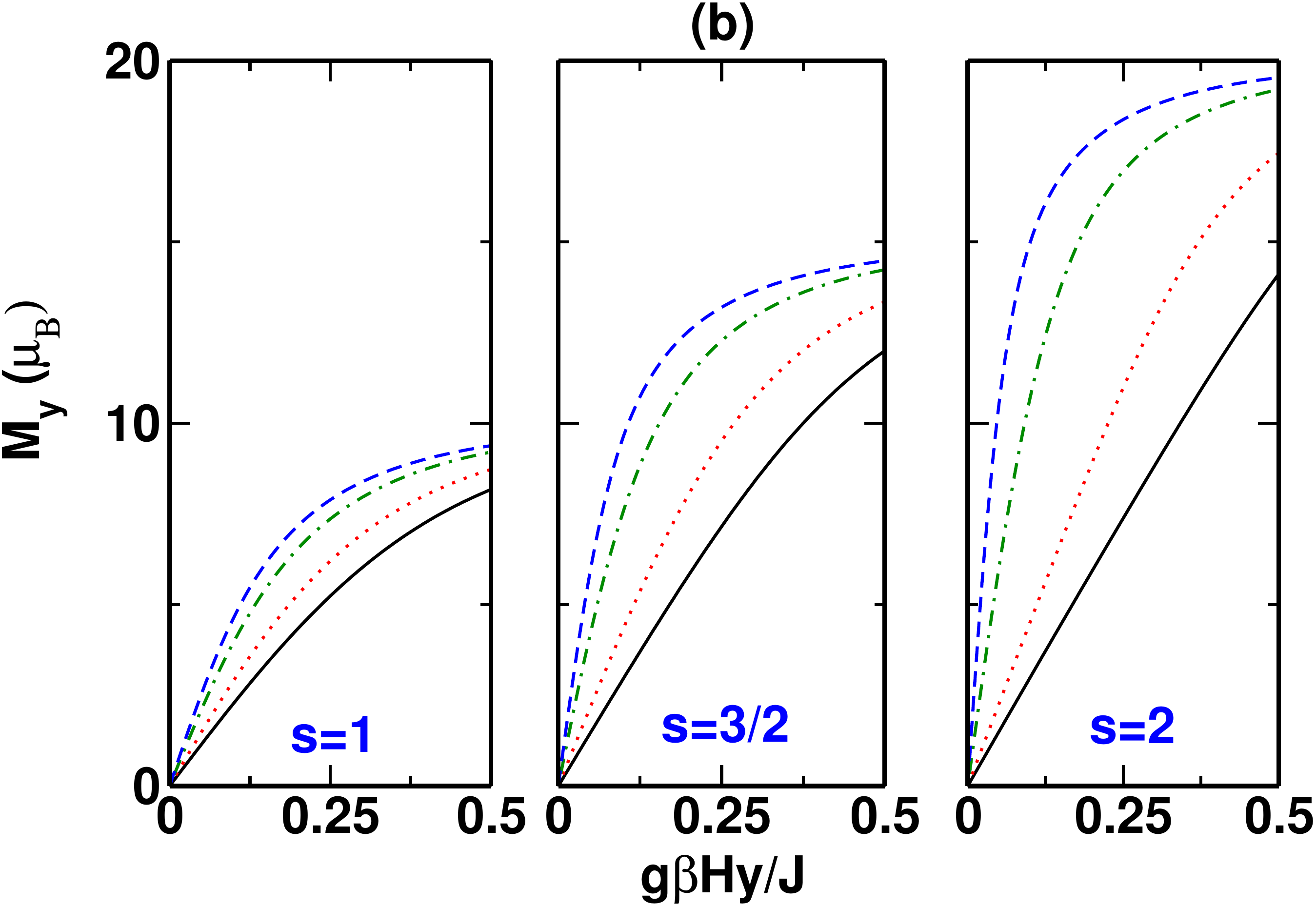}  
\center
      \includegraphics[width=8cm]{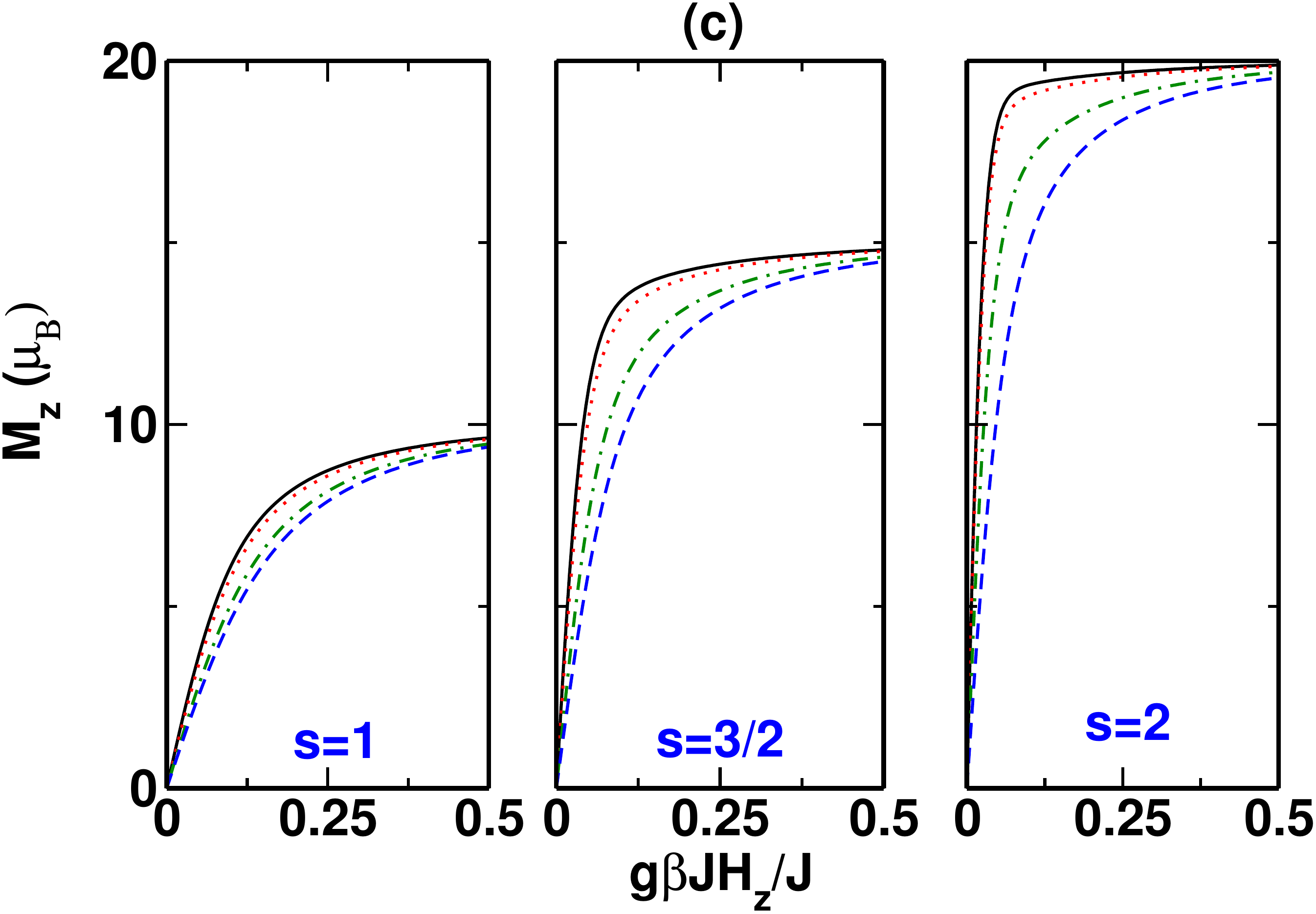} 
      \caption{\label{fig:MxHx,MyHy,MzHz} Dependence of magnetization ($M$) on applied magnetic field ($g\beta H/J$) at temperature $k_BT/J=1.0$. (a) $M_x$ versus $g\beta H_x/J$, (b) $M_y$ versus $g\beta H_y/J$ and (c) $M_z$ versus $g\beta H_z/J$ for different values of $\epsilon$ and $\delta$, in the absence of on-site anisotropy. Color coding and line type is the same for all panels.}
\end{figure*}
The specific heat behaviour is similar to the weak anisotropy case, we find a sharp peak at low temperature followed by a broad peak at higher temperatures. At higher anisotropies, we find a single peak in the $C_v$ vs $T$ plot \ref{fig:Cvxyz} and the temperature of the peak maxima is higher for higher anisotropy. For a fixed anisotropy, the peak maximum shifts to higher temperature as the site spin increases from $s=1$ to $s=2$.

\section{\label{sec:onsiteans}Systems with Exchange and On-Site Anisotropies}
In an earlier paper we discussed the role of on-site single ion anisotropy on the anisotropy of a spin chain. In this section we will discussed the effect of both exchange and on-site anisotropy on the magnetic properties of a spin chain  \citep{Haldar2017}.

We have introduced on-site anisotropy ($d/J$) in the eqn. \ref{eqn:generalHaml} and studied the spin chains with site spins $s=1$, $3/2$ and $2$ of length of five spins. We have also set $\delta=0$ and study only XXZ models in the presence of site anisotropy. We have taken same on-site anisotropy aligned along the z-axis for all the spins. When the on-site anisotropy is weak, we find that the resultant molecular magnetic anisotropy is nearly a sum of the molecular anisotropy due to on-site anisotropy alone and the molecular anisotropy due to exchange anisotropy alone. Thus, the two anisotropies are additive as seen in fig. \ref{fig:DMvsepsln}. This is true up to $\epsilon=0.1$ for all the site spins.

In table \ref{tab:xxztablewithd}, 
\begin{table}
\caption{\label{tab:xxztablewithd}Energy gaps (in units of $J$) from the ground state of the low-lying states lying below the lowest state with $M_s$=0 for $d/J=0.1$. $M_s$ is conserved and is a good quantum number. The total spin $S_{tot}$ is calculated from the expectation value of $\langle \hat S^2 \rangle$ of the state. Intruder states are shown in red.}
~~\\
\setlength{\tabcolsep}{1.0pt}
\begin{tabular}{|c|c|c|c|c|c|c|c|c|c|c|c|} \hline  
  & \multicolumn{11}{|c|} {N=5, d/J=0.1, XXZ model}   \\
 \cline{2-12}
 $\epsilon$ & \multicolumn{3}{|c|} {s=1} & &
 \multicolumn{3}{|c|} {s=3/2} & & 
 \multicolumn{3}{|c|} {s=2} \\
\cline{2-4} \cline{6-8} \cline{10-12}
& $M_s$ & $S_{tot}$ & Energy && $M_s$ & $S_{tot}$ & Energy && $M_s$ & $S_{tot}$ & Energy \\ \hline
~ 0.1 ~&~ $\pm 5$ ~&~ 5.00 ~&~ 0 ~&&~ $\pm 7.5$ ~&~ 7.50 ~&~ 0 ~&&~ $\pm 10$ ~&~ 10.00 ~&~ 0 ~\\
       &  $\pm 4$  &  4.99  &  0.412  &&  $\pm 6.5$  &  7.49  &  0.668  &&  $\pm 9$ &  9.99  &  0.9240 \\
       &\textcolor{red}{$\pm 4$}&\textcolor{red}{4.00}&\textcolor{red}{0.657}&&\textcolor{red}{$\pm 6.5$}&\textcolor{red}{6.50}&\textcolor{red}{1.035}&&\textcolor{red}{$\pm 9$}&\textcolor{red}{9.00}&\textcolor{red}{1.413} \\
       &  $\pm 3$  &  4.96  & 0.726 &&  $\pm 5.5$  &  7.47  & 1.238  &&  $\pm 8$ &  9.98  & 1.749 \\
       &\textcolor{red}{$\pm 3$}&\textcolor{red}{3.98}&\textcolor{red}{0.921}&&\textcolor{red}{$\pm 5.5$}&\textcolor{red}{6.48}&\textcolor{red}{1.551}&&\textcolor{red}{$\pm 8$}&\textcolor{red}{8.98}&\textcolor{red}{2.183} \\
       &  $\pm 2$  &  4.88  & 0.944 &&  $\pm 4.5$  &  7.43  &  1.708  &&  $\pm 7$ &  9.95  &  2.476 \\
       &  $\pm 1$  &  4.78  &  1.067  &&\textcolor{red}{$\pm 4.5$}&\textcolor{red}{6.44}&\textcolor{red}{1.969}&&\textcolor{red}{$\pm 8$}&\textcolor{red}{8.01}&\textcolor{red}{2.754} \\
       &\textcolor{red}{$\pm 2$}&\textcolor{red}{3.92}&\textcolor{red}{1.093}&&\textcolor{red}{$\pm 5.5$}&\textcolor{red}{5.53}&\textcolor{red}{1.998}&&\textcolor{red}{$\pm 7$}&\textcolor{red}{8.96}&\textcolor{red}{2.854} \\
       &     0   &  4.75  &  1.108  &&  $\pm 3.5$  &  7.36  &  2.080   &&  $\pm 6$ &  9.91  & 3.103  \\
       &         &         &           &&\textcolor{red}{$\pm 3.5$}&\textcolor{red}{6.36}&\textcolor{red}{2.291}&&\textcolor{red}{$\pm 9$}&\textcolor{red}{9.01}&\textcolor{red}{3.116} \\
       &         &         &           &&\textcolor{red}{$\pm 6.5$}&\textcolor{red}{6.51}&\textcolor{red}{2.312}&&\textcolor{red}{$\pm 7$}&\textcolor{red}{8.02}&\textcolor{red}{3.378} \\
       &         &         &           &&  $\pm 2.5$  &  7.26  & 2.351   &&  $\pm 6$ &  8.90  &  3.426 \\
       &         &         &           &&\textcolor{red}{$\pm 4.5$}&\textcolor{red}{5.56}&\textcolor{red}{2.373}&&  $\pm 5$ &  9.84  & 3.628 \\
       &         &         &           &&\textcolor{red}{$\pm 2.5$}&\textcolor{red}{6.26}&\textcolor{red}{2.520}&&\textcolor{red}{$\pm 5$}&\textcolor{red}{8.82}&\textcolor{red}{3.900} \\
       &         &         &           &&  $\pm 1.5$  &  7.15  &  2.5270   &&\textcolor{red}{$\pm 6$}&\textcolor{red}{8.03}&\textcolor{red}{3.907} \\
       &         &         &           &&  $\pm 0.5$  &  7.08  &  2.6130   &&\textcolor{red}{$\pm 8$}&\textcolor{red}{8.99}&\textcolor{red}{3.908} \\
       &         &         &           &&             &         &            &&\textcolor{red}{$\pm 7$}&\textcolor{red}{7.04}&\textcolor{red}{4.023} \\
       &         &         &           &&             &         &            &&  $\pm 4$ &  9.75  &  4.053\\
       &         &         &           &&             &         &            &&\textcolor{red}{$\pm 4$}&\textcolor{red}{8.71}&\textcolor{red}{4.276} \\
       &         &         &           &&             &         &            &&\textcolor{red}{$\pm 5$}&\textcolor{red}{8.01}&\textcolor{red}{4.343} \\
       &         &         &           &&             &         &            &&\textcolor{red}{$\pm 8$}&\textcolor{red}{8.01}&\textcolor{red}{4.366} \\
       &         &         &           &&             &         &            &&  $\pm 3$ &  9.63   &  4.377 \\
       &         &         &           &&             &         &            &&\textcolor{red}{$\pm 6$}&\textcolor{red}{7.08}&\textcolor{red}{4.503} \\
       &         &         &           &&             &         &            &&\textcolor{red}{$\pm 3$}&\textcolor{red}{8.58}&\textcolor{red}{4.560} \\
       &         &         &           &&             &         &            &&  $\pm 2$ &  9.51  & 4.604  \\
       &         &         &           &&             &         &            &&\textcolor{red}{$\pm 7$}&\textcolor{red}{8.98}&\textcolor{red}{4.605} \\
       &         &         &           &&             &         &            &&\textcolor{red}{$\pm 4$}&\textcolor{red}{8.01}&\textcolor{red}{4.688} \\
       &         &         &           &&             &         &            &&  $\pm 1$ &  9.41  &  4.737 \\
       &         &         &           &&             &         &            &&\textcolor{red}{$\pm 2$}&\textcolor{red}{8.47}&\textcolor{red}{4.755} \\
       &         &         &           &&             &         &            &&     0  &  9.37  &  4.781 \\ \hline
\end{tabular}
\end{table}
we show the low-energy spectrum of the $N=5$ spin chain for $s=1$, $3/2$ and $2$, where both exchange and on-site anisotropies are large. In cases where we can not define the molecular magnetic anisotropy in terms of the parameter $D_M$ of the effective spin Hamiltonian, we follow the system by computing the magnetic susceptibilities, magnetization and specific heat. We have shown in fig. \ref{fig:ChiXXd,ChiZZd}, the difference in the $\Delta\chi_{_{xx}}T~=~\chi_{_{xx}}T(\epsilon, d \neq 0)-\chi_{_{xx}}T(\epsilon, d=0)$ and $\Delta\chi_{_{zz}}T~=~\chi_{_{zz}}T(\epsilon, d \neq 0)-\chi_{_{zz}}T(\epsilon, d=0)$ of magnetic susceptibility as a function of $d/J$ at $k_BT/J=1$ for different $\epsilon$ values. We find that nonzero $d$ enhances $\Delta\chi_{_{zz}}T$ but decreases $\Delta\chi_{_{xx}}T$ values. In case of site spin $s=1$, the dependence of $\Delta\chi_{_{xx}}T$ and $\Delta\chi_{_{zz}}T$ on site anisotropy is weak and linear. In case of $s=3/2$ and $s=2$ the difference $\Delta\chi_{_{zz}}T$ increases sharply as $d/J$ is increased and for higher $d/J$ it tends to saturate. The saturation is more apparent in the $s=2$ case. $\Delta\chi_{_{xx}}T$ on the other hand decreases with increasing $d/J$. This is because the on-site anisotropy is oriented along the z-axis. This is also the reason why $\Delta\chi_{_{xx}}T$ shows a sharper drop with $d/J$ for larger $\epsilon$ while $\Delta\chi_{_{zz}}T$ shows a sharper rise for larger $\epsilon$.
\begin{figure}
    \includegraphics[width=10cm]{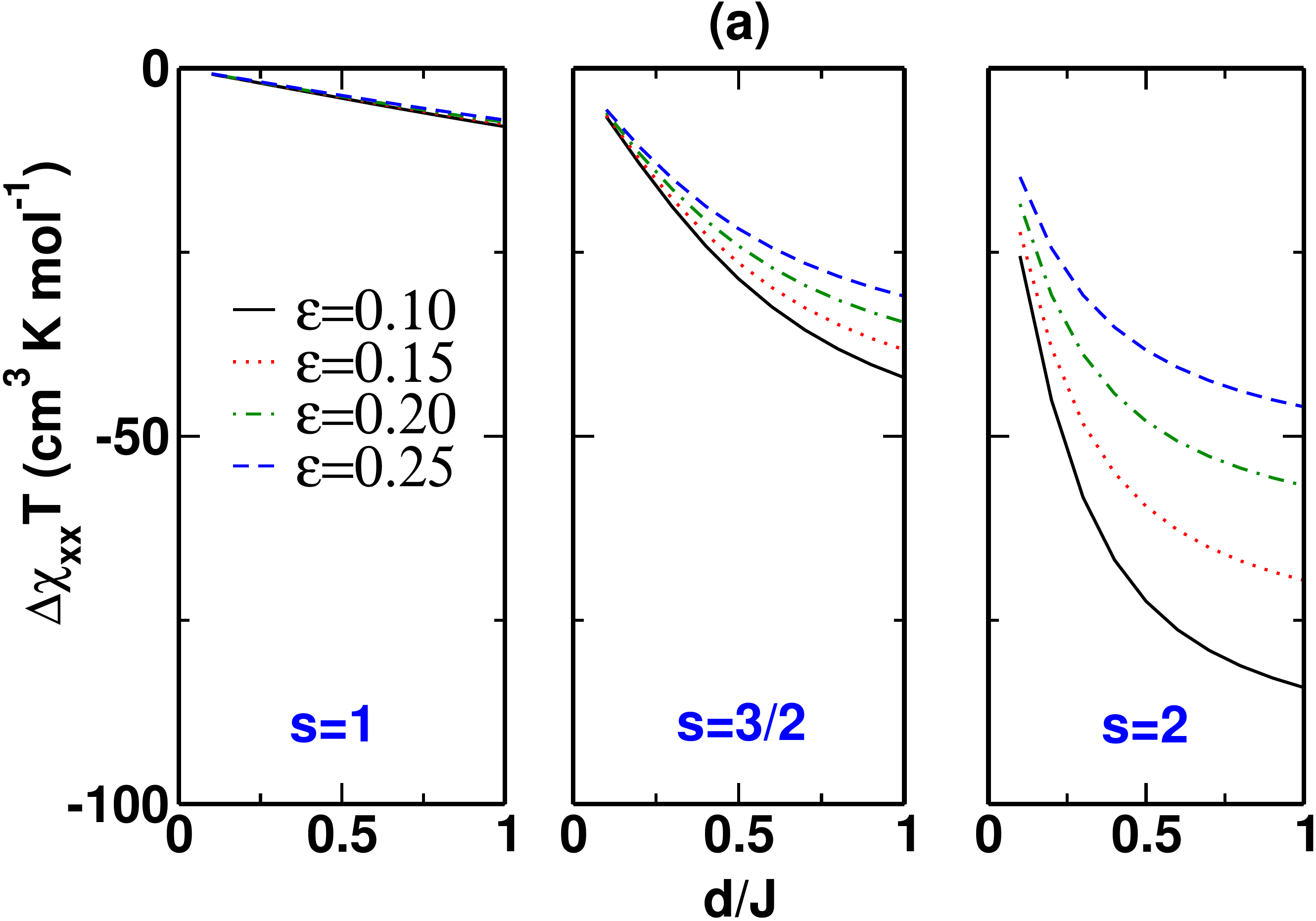}  

\vspace{0.3in}
\includegraphics[width=10cm]{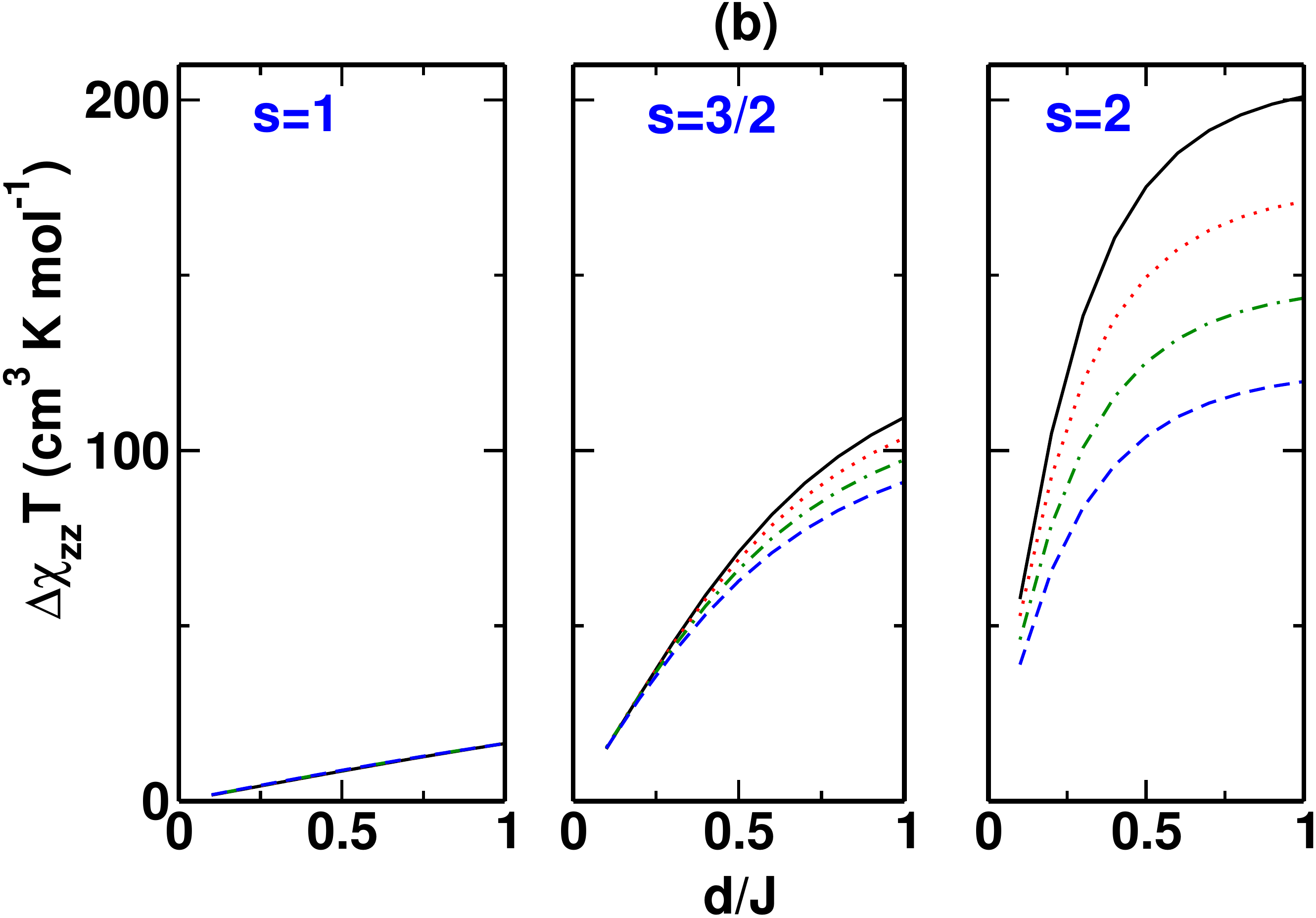} 
      \caption{\label{fig:ChiXXd,ChiZZd}The effect of on-site anisotropy $d/J$ on (a)$\Delta\chi_{_{xx}}T=[\chi_{_{xx}}T(\epsilon, d \neq 0)-\chi_{_{xx}}T(\epsilon, d=0)]$ at $g\beta H_x/J=0.005$, $k_BT/J=1.0$ and (b) $\Delta\chi_{_{zz}}T=[\chi_{_{zz}}T(\epsilon, d \neq 0)-\chi_{_{zz}}T(\epsilon, d=0)]$ at $g\beta H_z/J=0.005$, $k_BT/J=1.0$ for $\epsilon=0.10$, $0.15$, $0.20$ and $0.25$. Same color code and line type is used for all panels. Also note the sign of $\Delta\chi_{_{xx}}T$ is -ve while $\Delta\chi_{_{zz}}T$ is +ve.  }
\end{figure}
Similarly in fig. \ref{fig:MxHxd,MzHzd}, we plot $\Delta M_x$ and $\Delta M_z$ for different $s$ and $\epsilon$, as a function of $d/J$. The field strength is $g\mu_BH~=~J/2$. We note that the $\Delta M_x$ decreases sharply with $d/J$ for $s=2$ and large $\epsilon$ while $\Delta M_z$ increases with $d/J$ and saturates for $s=2$ case while in the $s=3/2$ and $s=1$ cases, the saturation does not occur even for $d/J=1.0$. Again $\Delta M_z$ is larger when $\epsilon$ is small while $\Delta M_x$ is larger for large $\epsilon$.

\begin{figure}
    \includegraphics[width=10cm]{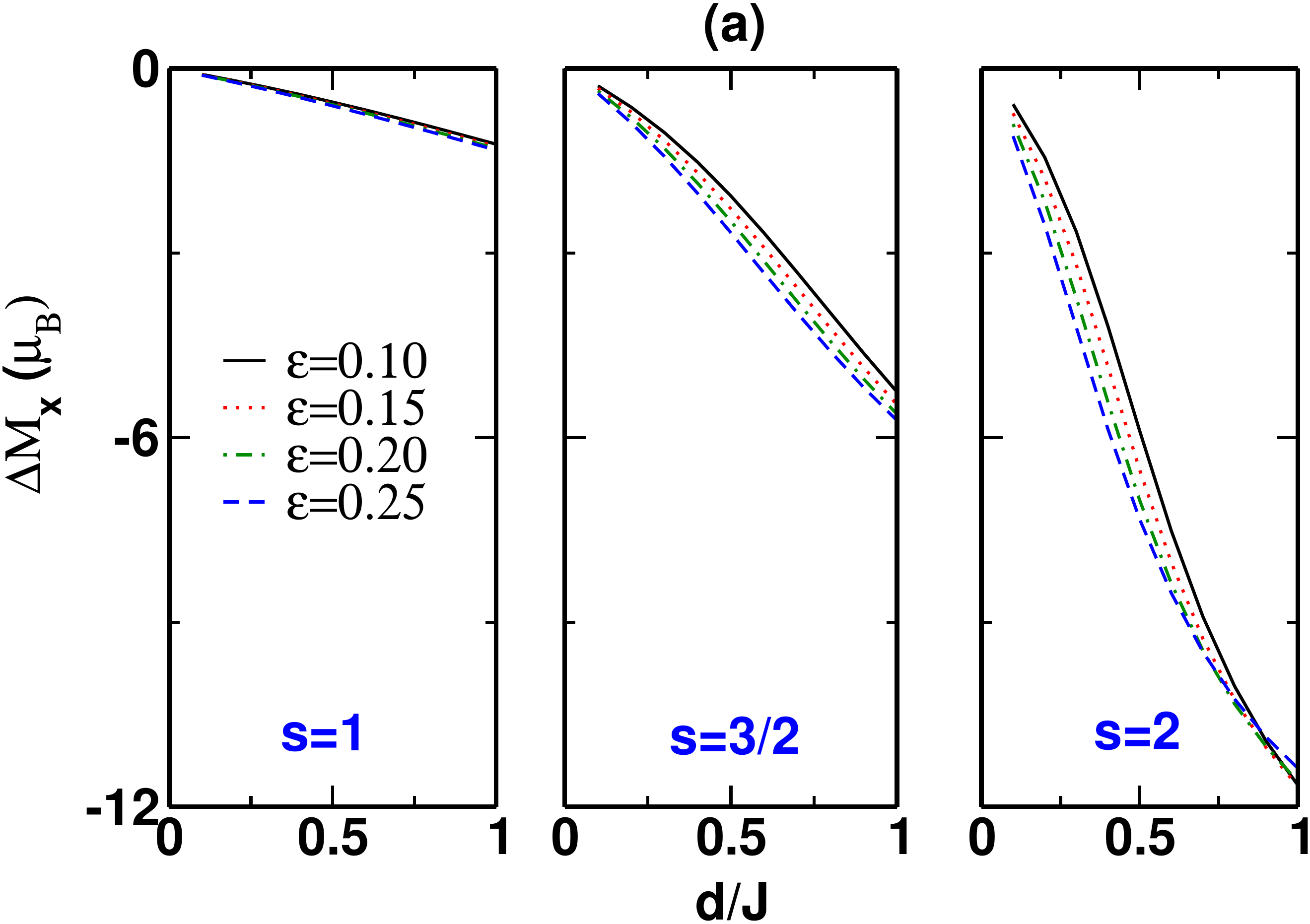}  

 \vspace{0.3in}
 \includegraphics[width=10cm]{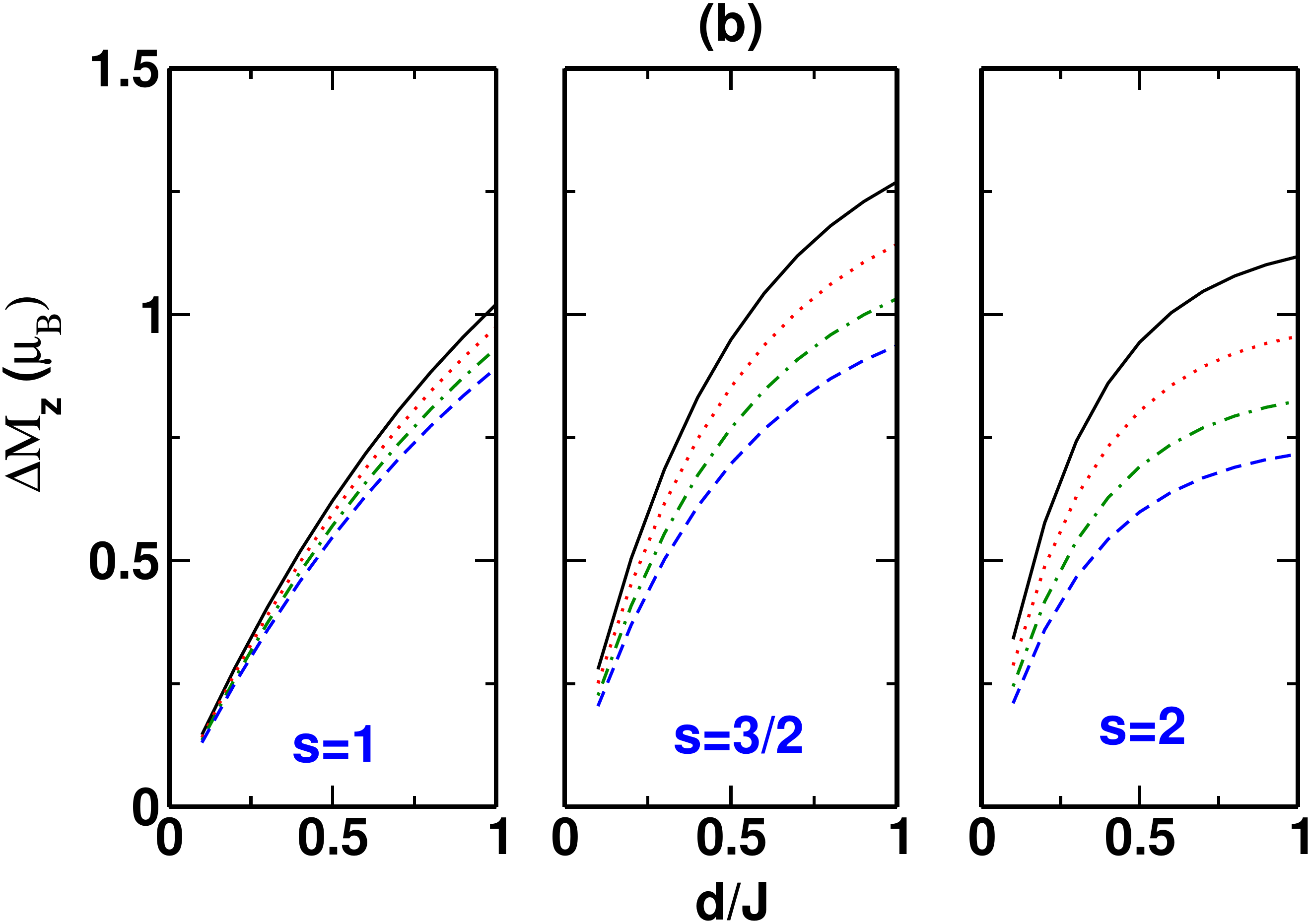} 
      \caption{\label{fig:MxHxd,MzHzd}The effect of on-site anisotropy $d/J$ on (a) $\Delta M_x=M_x(\epsilon, d \neq 0)-M_x(\epsilon, d=0)$ at $g\beta H_x/J=0.25$, $k_BT/J=1.0$ and (b) $\Delta M_z=M_z(\epsilon, d \neq 0)-M_z(\epsilon, d=0)$ at $g\beta H_z/J=0.25$, $k_BT/J=1.0$ for $\epsilon=0.10$, $0.15$, $0.20$ and $0.25$. Same color and line type is used for all panels.}
\end{figure}

\begin{figure}
\center
    \includegraphics[width=12cm]{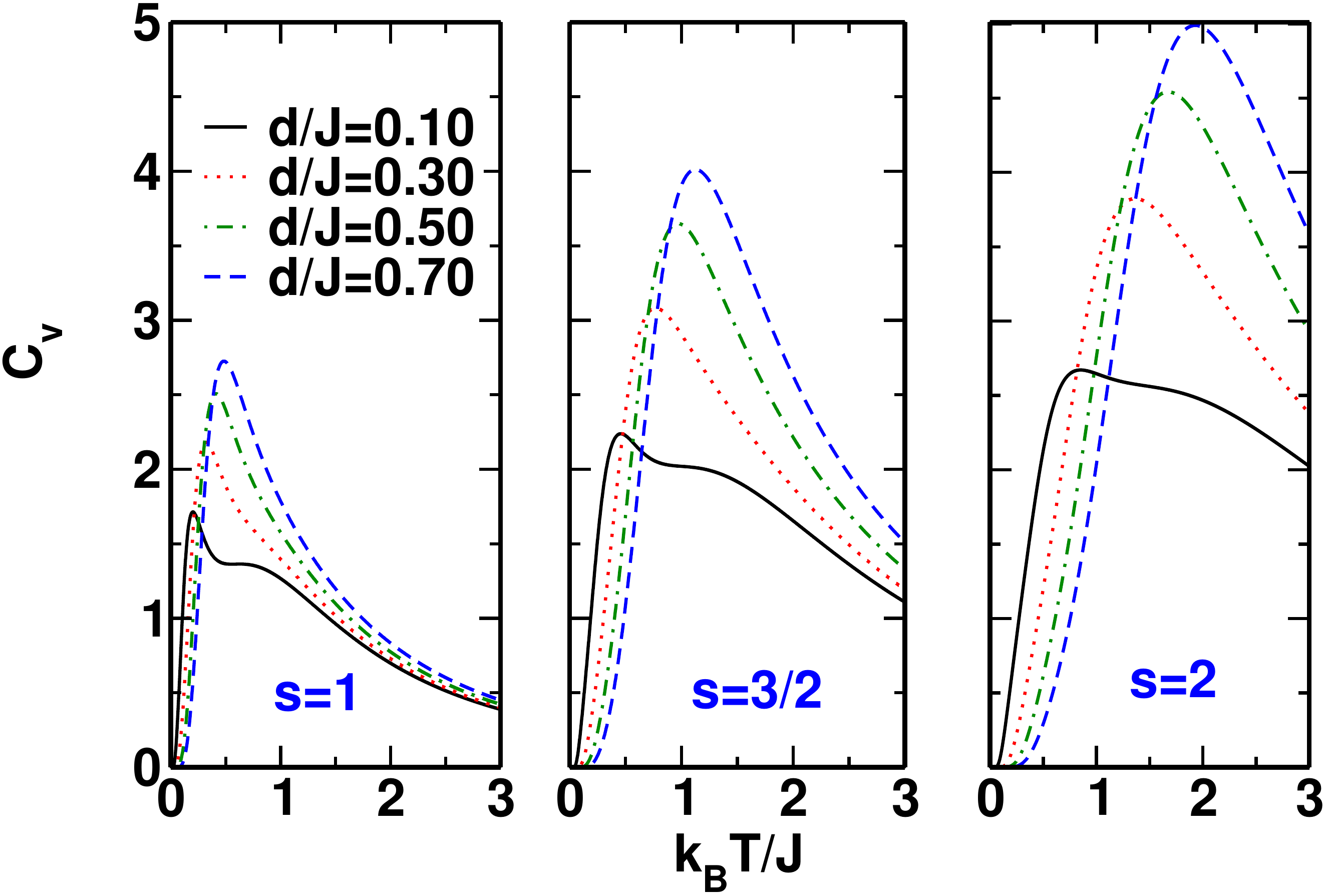} 
      \caption{\label{fig:Cvxxzd}Dependence of specific heat ($C_v$) on temperature ($k_BT/J$) of spin chains with $s=1$, $3/2$ and $s=2$ with systems size $N=5$ for $\epsilon=0.10$ in the presence of $d/J=0.10$, $0.30$, $0.50$ and $0.70$. Same color and line type is used for all panels. }
\end{figure}

The specific heat behaviour is shown in fig. \ref{fig:Cvxxzd}. We find that the two peak structure persists for small $d/J$ for $\epsilon=0.1$. However, increasing $d/J$ leads to a single peak. The peak position shifts to higher temperatures as $d/J$ increases and the peak also becomes sharper as $d/J$ increases. This is true for all site spins.

\newpage
\section{\label{sec:conclusion}Conclusions}
Our study of anisotropic ferromagnetic exchange models with site anisotropy shows that for small exchange and site anisotropies, the energy level splitting of the total spin states can be characterized by the axial anisotropy parameter $D_M$ which is a sum of the exchange alone  and ion anisotropy alone $D_M$ parameters. For large anisotropic exchange, neither the total spin nor its z-component are conserved and it is not possible to define the molecular anisotropy parameters $D_M$ and $E_M$. The effect of anisotropy is then studied by following thermodynamics properties such as $\chi$, $C_v$ and $M$. This is also true when the on-site anisotropy is large, even in the absence of exchange anisotropy. We find two peak structure in $C_v$ vs $T$ when the exchange is weakly anisotropic. We also find that this feature prevails for weak on-site anisotropy as well. The dual peak structure is more pronounced for smaller on-site spins. In general the effect of anisotropy, as seen form the presence of intruder states from different parent spin state, is more pronounced in the case of higher site spins and longer chain length. The synergy between site anisotropy and exchange anisotropy becomes complicated when both are strong. We observe that the difference in susceptibilities as well as magnetization as a function of the site anisotropy strength for large exchange anisotropy becomes highly nonlinear, particularly for systems with higher site spin.
\section{\label{sec:ack}Acknowledgements}
\noindent SR and JPS acknowledge the support through IFCPAR/CEFIPRA projets. SR also thanks DST for support through different projects and a fellowship and Indian Science Academy for Senior Scientist Position. RR thanks TUE-DST for support.


\end{document}


\title{Modeling Molecular Magnets with Large Exchange and On-Site Anisotropies}

\author[a]{Sumit Haldar\thanks{E-mail: sumithaldar@iisc.ac.in}}
\author[b]{Rajamani Raghunathan\thanks{E-mail: rajamani@csr.res.in}}
\author[c]{Jean-Pascal Sutter\thanks{E-mail: jean-pascal.sutter@lcc-toulouse.fr}} 
\author[a]{S. Ramasesha\thanks{E-mail: ramasesh@iisc.ac.in}}
\affil[a]{Solid State and Structural Chemistry Unit, Indian Institute of Science, Bangalore - 560012, India.}
\affil[b]{UGC-DAE Consortium for Scientific Research, Indore - 452017, India.}
\affil[c]{LCC-CNRS, Universit\'e de Toulouse, UPS, INPT, CNRS, Toulouse, France.}

\maketitle

\section{Corresponding Results for Spin Chains with 4 spins:}
\begin{table}
\caption{\label{tab:XXZTableFour}Energy gaps from the ground state of the low-lying states within the lowest state with $M_s$=0. $M_s$ is conserved and is a good quantum number. The total spin $S$ is calculated from the expectation value of $\langle \hat S^2 \rangle$ of the state. Intruder states are shown in red.}
~~\\
\setlength{\tabcolsep}{1.0pt}
\begin{tabular}{|c|c|c|c|c|c|c|c|c|c|c|c|} \hline  
  & \multicolumn{11}{|c|} {N=4, XXZ model}   \\
 \cline{2-12}
 $\epsilon$ & \multicolumn{3}{|c|} {s=1} & &
 \multicolumn{3}{|c|} {s=3/2} & & 
 \multicolumn{3}{|c|} {s=2} \\
\cline{2-4} \cline{6-8} \cline{10-12}
 &$M_s$ & $S_{tot}$& Energy && $M_s$ & $S_{tot}$ & Energy && $M_s$ & $S_{tot}$ & Energy\\ \hline
~ 0.10 ~&~ $\pm 4$ ~&~ 4.00 ~&~ 0 ~&&~ $\pm 6$ ~&~ 6.00 ~&~ 0 ~&&~ $\pm 8$ ~&~ 8.00 ~&~ 0~ \\
        &~ $\pm 3$ ~&~ 3.99 ~&~0.149  ~&&~ $\pm 5$ ~&~ 5.99 ~&~ 0.223 ~&&~ $\pm 7$ ~&~ 7.99 ~&~ 0.297~ \\
        &~ $\pm 2$ ~&~ 3.99 ~&~ 0.255 ~&&~ $\pm 4$ ~&~ 5.99 ~&~ 0.406 ~&&~ $\pm 6$ ~&~ 7.99 ~&~ 0.556~ \\
        &~ $\pm 1$ ~&~ 3.99 ~&~ 0.320 ~&&~ $\pm 3$ ~&~ 5.99 ~&~ 0.549 ~&&~ $\pm 5$ ~&~ 7.99 ~&~ 0.776~ \\
        &     0   &~ 3.99 ~&~  0.342  ~&&~ $\pm 2$~&~ 5.99 ~&~ 0.653 ~&&~  $\pm 4$ ~&~ 7.99 ~&~ 0.956~ \\
        &           &    &              &&~ $\pm 1$ ~&~ 5.99 ~&~ 0.714 ~&&~ $\pm 3$ ~&~ 7.99 ~&~ 1.096~  \\
        &           &    &              &&     0  ~&~ 5.99 ~&~ 0.734 ~&&~ $\pm 2$ ~&~ 7.99 ~&~1.197 ~ \\
        &           &    &              &&          &    &               &&~ $\pm 1$ ~&~ 7.99 ~&~ 1.257~ \\
        &           &    &              &&          &    &               &&~    0  ~&~ 7.99 ~&~ 1.277~ \\ \hline
\hline
0.20 & $\pm 4$ & 4.00 &~0~ &&$\pm 6$ & 6.00 & 0 && $\pm 8$ & 8.00 & 0\\
     & $\pm 3$ & 3.99 &~0.294~ &&$\pm 5$ & 5.99 & 0.441 && $\pm 7$ & 7.99 & 0.587\\
     & $\pm 2$ & 3.99 &~0.507~ &&$\pm 4$ & 5.99 & 0.806 && $\pm 6$ & 7.99 & 1.102\\
     & $\pm 1$ & 3.99 &~0.635~ &&\textcolor{red}{$\pm 5$}&\textcolor{red}{5.00}&\textcolor{red}{1.044}&&\textcolor{red}{$\pm 7$} &\textcolor{red}{7.00}&\textcolor{red}{1.392}\\
     &    0  & 3.99 &~0.678~ &&$\pm 3$ & 5.99 &  1.092 && $\pm 5$ & 7.99 &1.541 \\
     &           &    &         &&$\pm 2$ & 5.99 & 1.298 && $\pm 4$ & 7.99 & 1.902\\
     &           &    &         &&\textcolor{red}{$\pm 4$}&\textcolor{red}{4.99}&\textcolor{red}{1.406}&&\textcolor{red}{$\pm 6$}&\textcolor{red}{6.99}&\textcolor{red}{1.908}\\
     &           &    &         &&$\pm 1$ & 5.99 & 1.422 && $\pm 3$ & 7.99 &2.184\\
     &           &    &         &&   0  & 5.99 & 1.463 &&\textcolor{red}{$\pm 5$}&\textcolor{red}{6.99}&\textcolor{red}{2.346}\\ 
     &           &    &         &&          &    &        && $\pm 2$ & 7.99 &2.386\\
     &           &    &         &&          &    &        && $\pm 1$ & 7.99 &2.507\\
     &           &    &         &&          &    &        &&    0  & 7.99 &2.548\\ \hline
\hline
0.25 & $\pm 4$ & 4.00&~0~&&$\pm 6$ & 6.00 &~0~&&$\pm 8$ & 8.00 &~ 0~ \\
     & $\pm 3$ & 3.99&~0.365~&&$\pm 5$ & 5.99 &~0.547~&&$\pm 7$ & 7.99 &~ 0.729~ \\
     & $\pm 2$ & 3.99&~0.631~&&$\pm 4$ & 5.99 &~1.002~&&$\pm 6$ & 7.99 &~ 1.370~ \\
     &\textcolor{red}{$\pm 3$}&\textcolor{red}{3.00}&~\textcolor{red}{0.723}~&&\textcolor{red}{$\pm 5$}&\textcolor{red}{5.00}&~\textcolor{red}{1.084}~&&\textcolor{red}{$\pm 7$}&\textcolor{red}{7.00}&\textcolor{red}{1.445} \\
     & $\pm 1$ & 3.98&~0.791~&&$\pm 3$ & 5.98 &~1.360~&&$\pm 5$ & 7.99 &~ 1.918~ \\
     &   0  & 3.98&~0.844~&&\textcolor{red}{$\pm 4$}&\textcolor{red}{5.00}&\textcolor{red}{1.539}&&\textcolor{red}{$\pm 6$}&\textcolor{red}{6.99}&\textcolor{red}{2.088} \\
     &          &    &       &&$\pm 2$ & 5.98 &~1.617~&&$\pm 4$ & 7.99 &~2.369 ~ \\
     &           &    &      &&$\pm 1$ & 5.98 &~1.773~&&\textcolor{red}{$\pm 5$}&\textcolor{red}{6.99}&\textcolor{red}{2.634} \\
     &           &    &      &&  0   & 5.98 &~1.824~&&$\pm 3$ & 7.98 &~ 2.723~ \\ 
     &           &    &       &&          &    &       &&\textcolor{red}{$\pm 6$}&\textcolor{red}{6.00}&\textcolor{red}{2.891} \\
     &           &    &       &&          &    &       &&$\pm 2$ & 7.98 &~ 2.976~ \\
     &           &    &       &&          &    &       &&\textcolor{red}{$\pm 4$}&\textcolor{red}{6.98}&\textcolor{red}{3.082} \\
     &          &     &       &&          &    &       &&$\pm 1$ & 7.98 &~3.129 ~ \\
     &          &     &       &&          &    &       &&  0   & 7.98 &~ 3.180~ \\ \hline
\end{tabular}
\end{table}

\begin{figure*}
    \includegraphics[width=10cm]{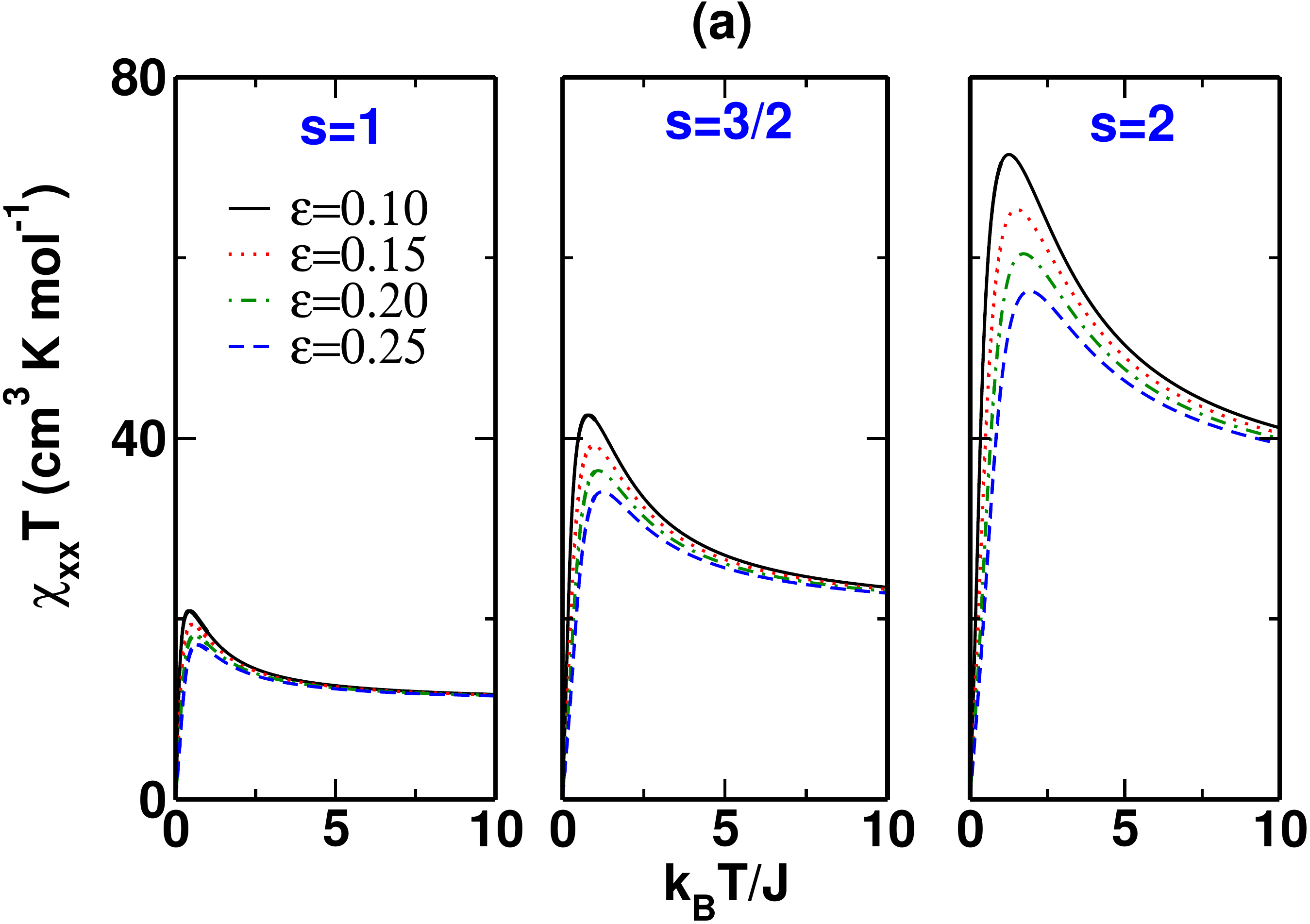}  
    
    \vspace{0.3in}
\includegraphics[width=10cm]{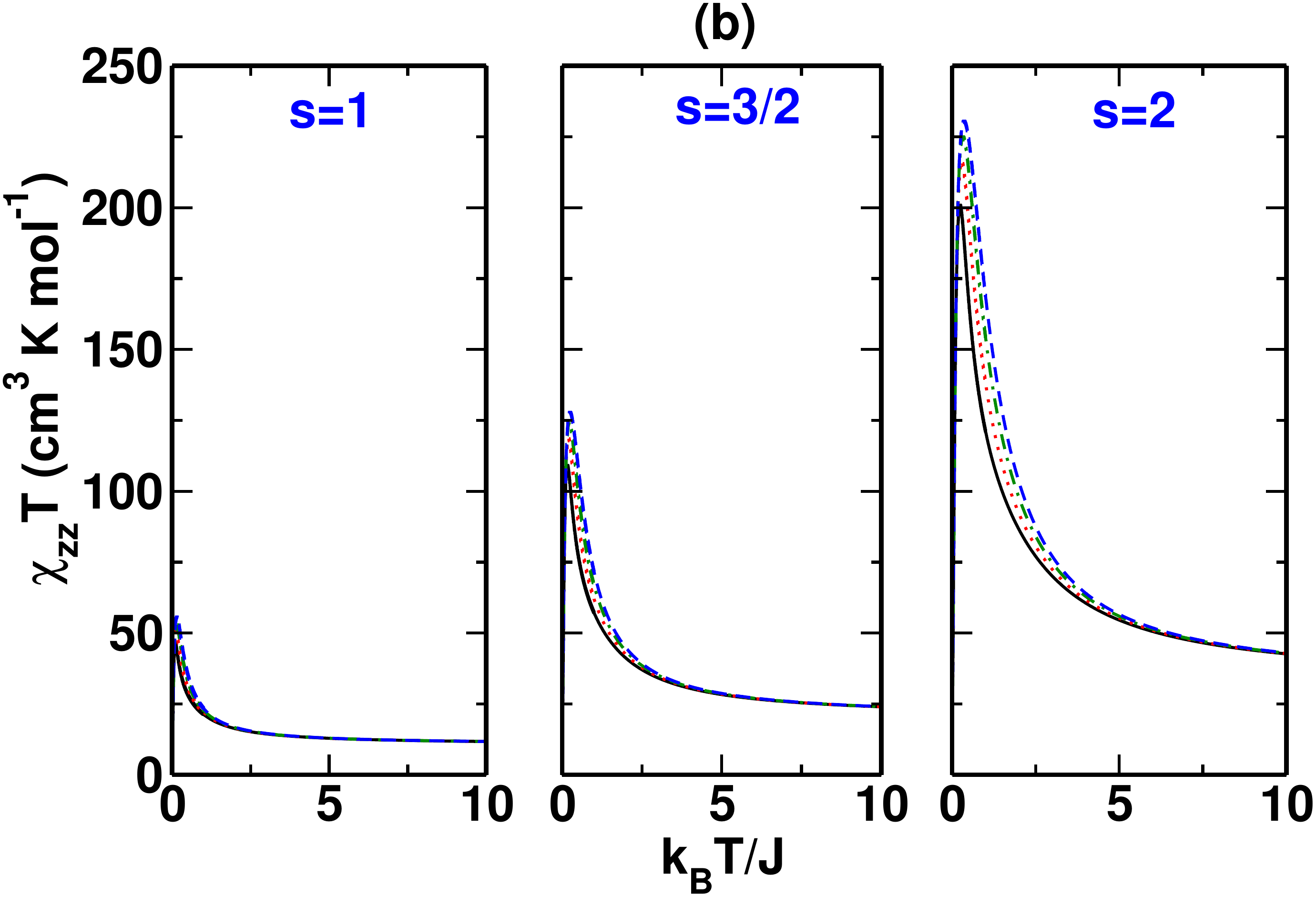} 
      \caption{\label{fig:ChiXX,ChiZZ}(a) Plot of $\chi_{_{xx}}T$ as a function of temperature computed with applied field along x-direction, (b) Plot of $\chi_{_{zz}}T$ as a function of temperature computed with applied field along z-direction for different values of exchange anisotropy $\epsilon$, in the absence of on-site anisotropy. The susceptibilities are computed for field magnitude $H=J/g\beta=0.005$. Color coding and line type is the same for all panels. (Note the scale on y-axis are different for (a) and (b)).}
\end{figure*}

\begin{figure*}
    \includegraphics[width=10cm]{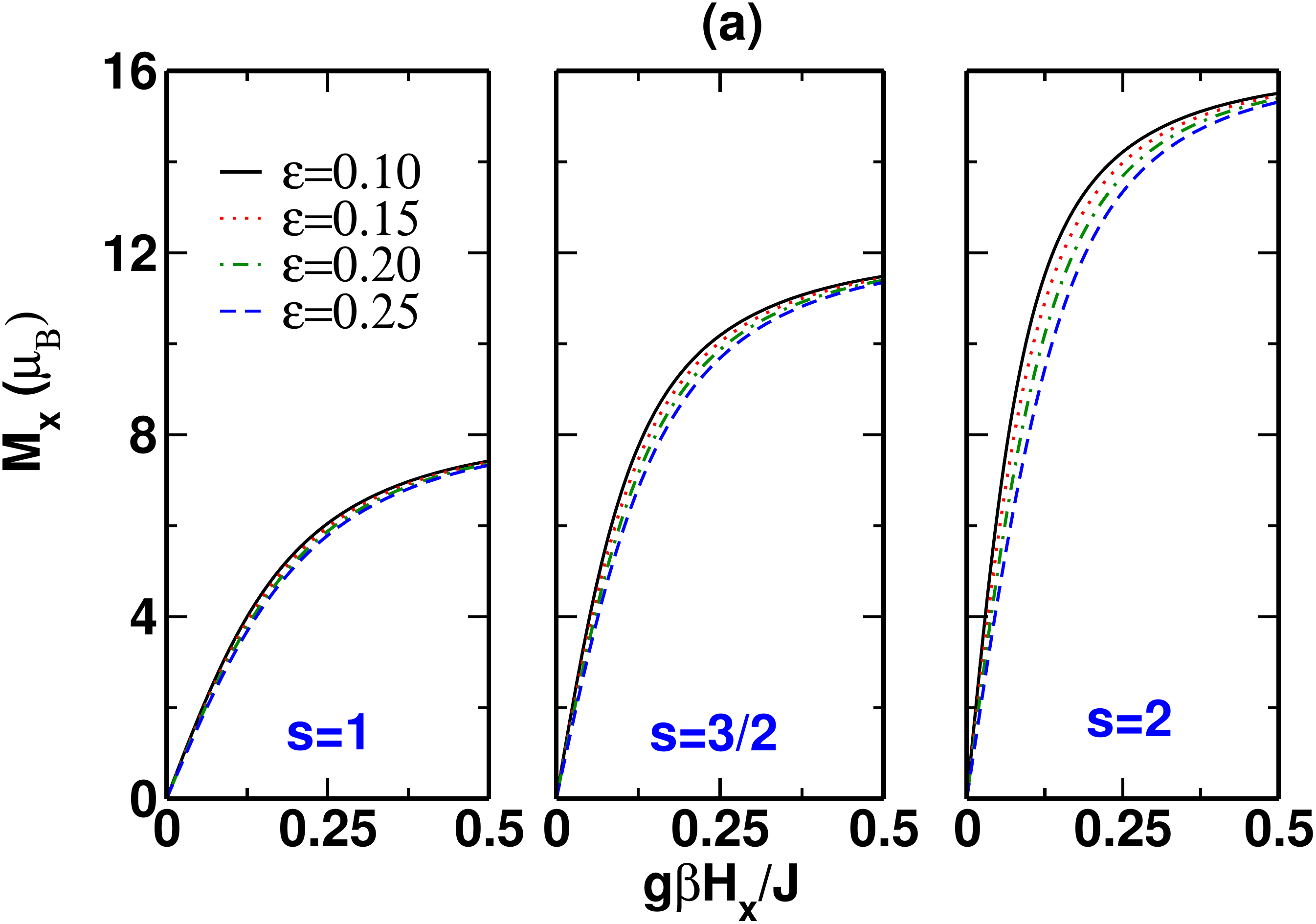}  

\vspace{0.3in}
\includegraphics[width=10cm]{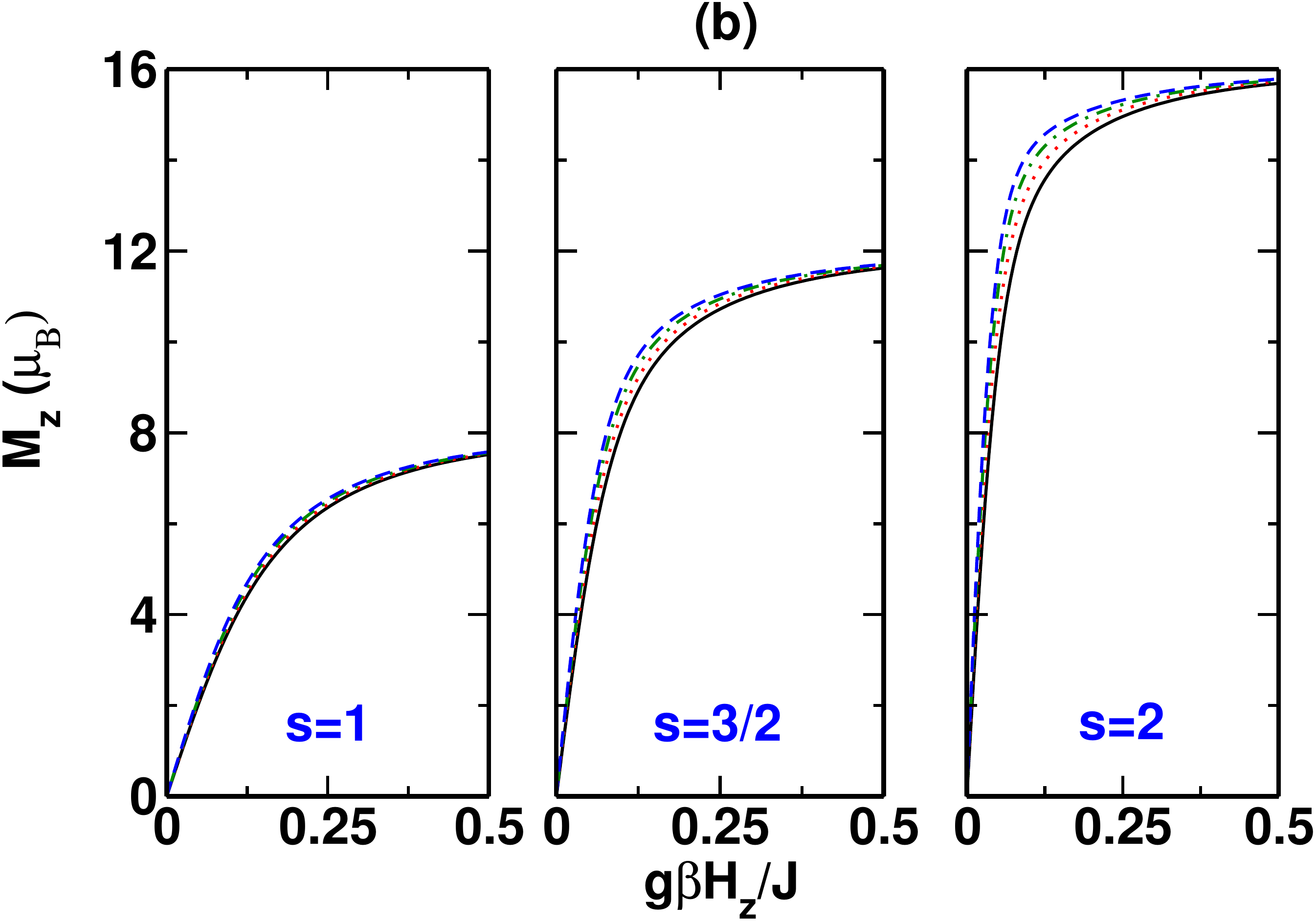} 
      \caption{\label{fig:MxHx,MzHz}Dependence of magnetization ($M$) on applied magnetic field ($g\beta H/J$) at temperature $k_BT/J=1.0$. (a) $M_x$ versus $g\beta H_x/J$, (b) $M_z$ versus $g\beta H_z/J$ for different values of exchange anisotropy $\epsilon$, in the absence of on-site anisotropy. Color coding and line type is the same for all panels.}
\end{figure*}

\begin{figure*}
    \includegraphics[width=12cm]{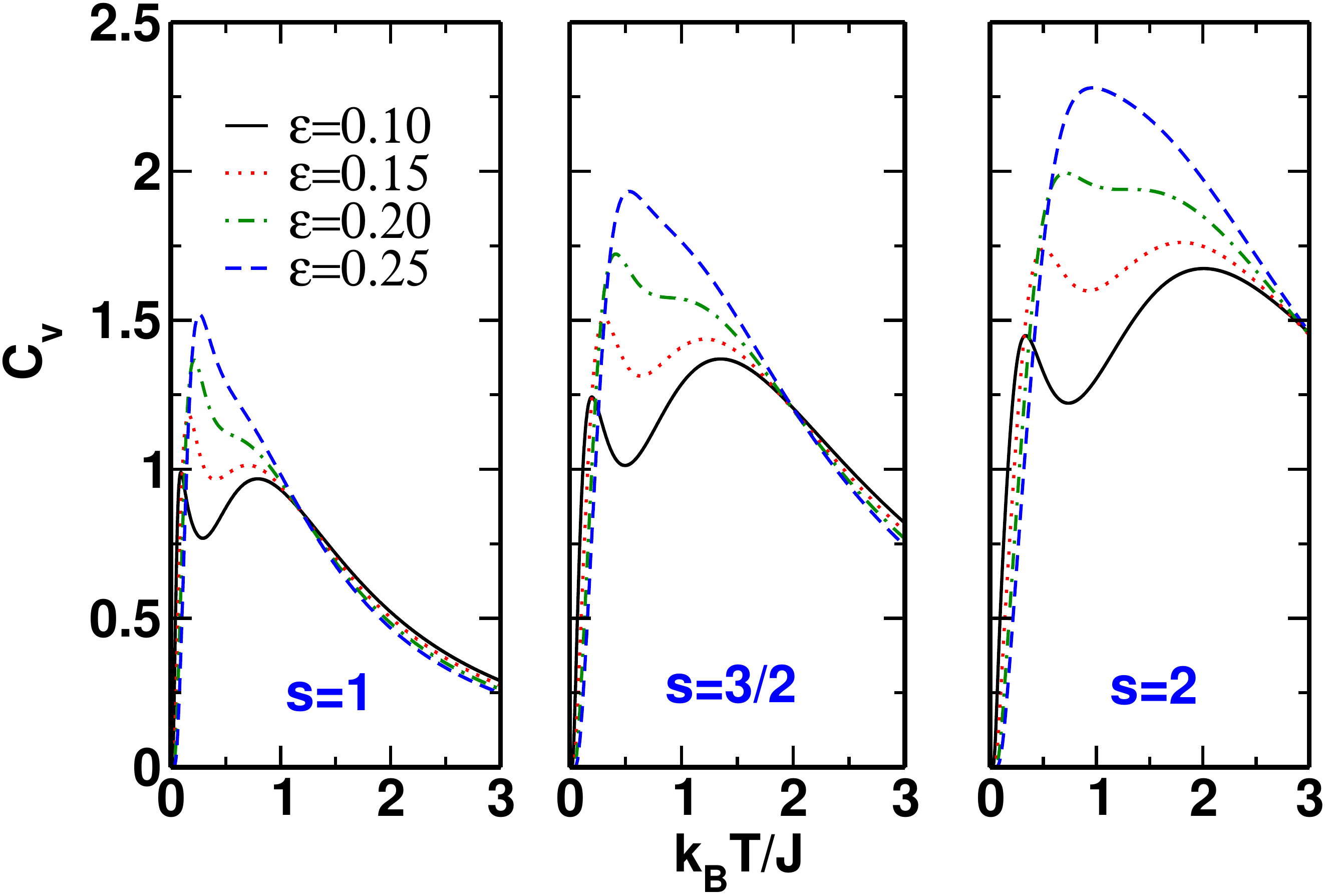} 
      \caption{\label{fig:Cvxxz}Dependence of specific heat ($C_v$) on temperature ($k_BT/J$) of spin chains with $s=1$, $3/2$ and $s=2$ for different values of axial exchange anisotropy $\epsilon$, in the absence of on-site anisotropy. Color coding and line type is the same for all panels.}
\end{figure*}

\begin{figure*}
\center
    \includegraphics[width=8cm]{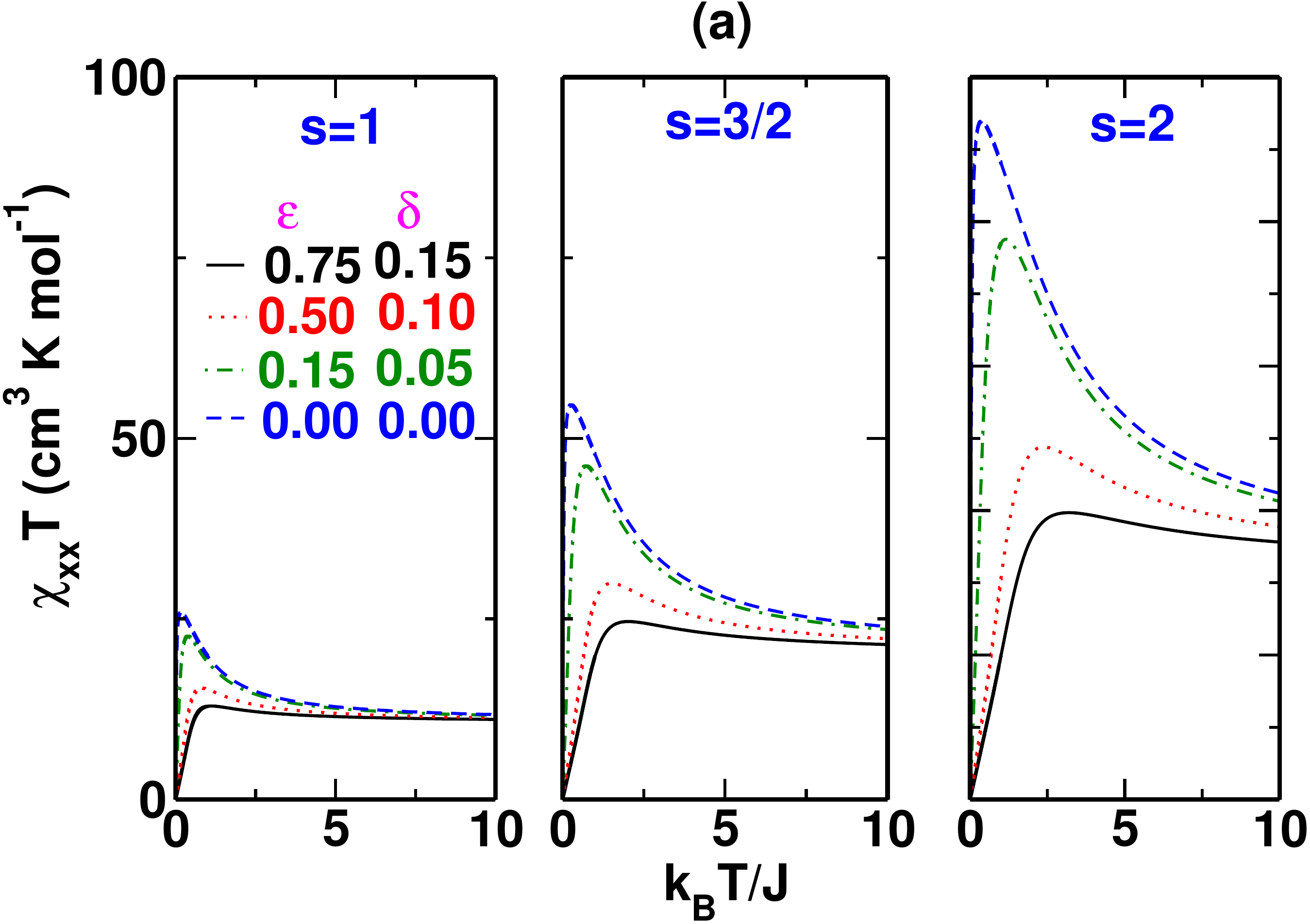}
 \center
     \includegraphics[width=8cm]{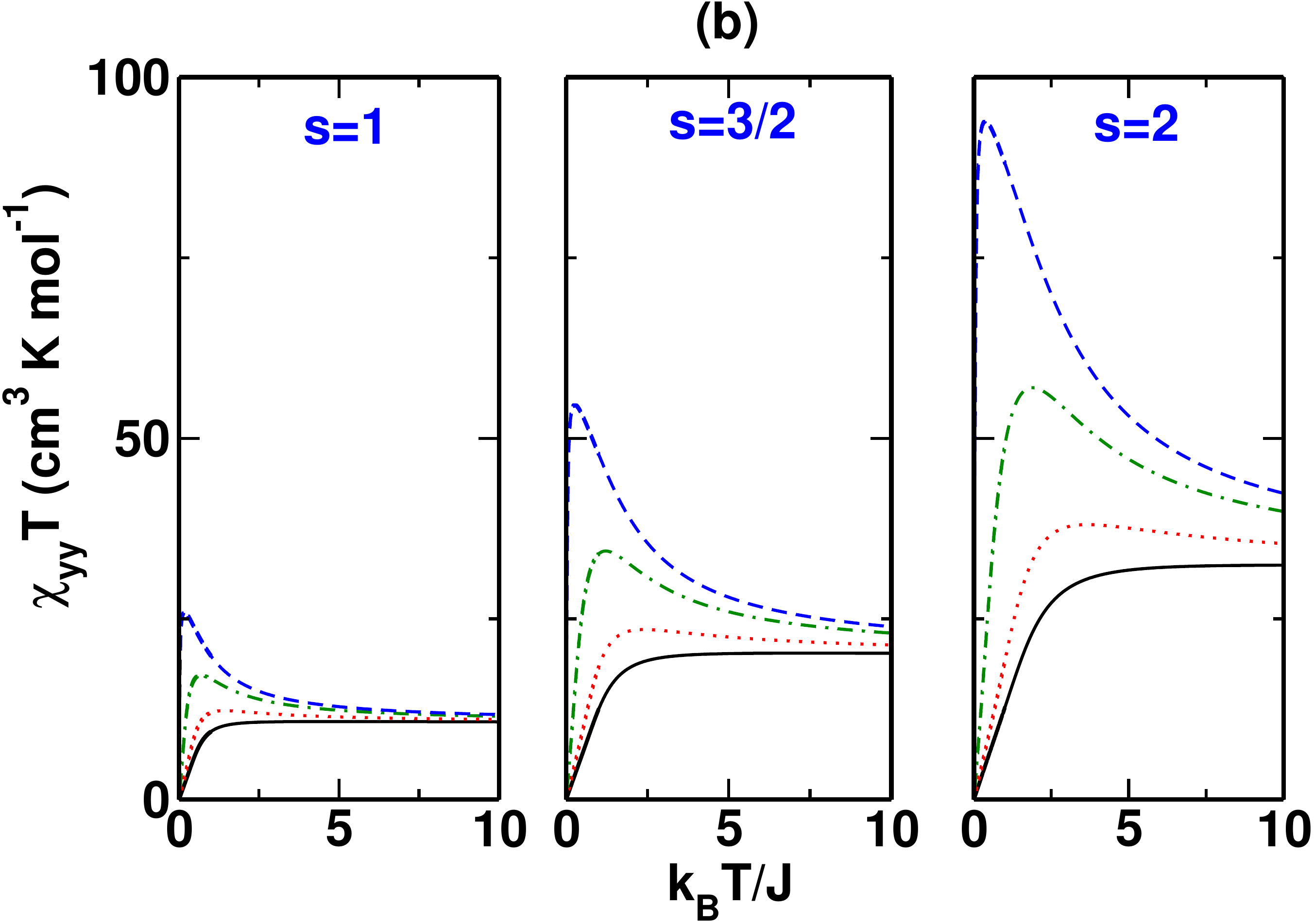}  
\center
     \includegraphics[width=8cm]{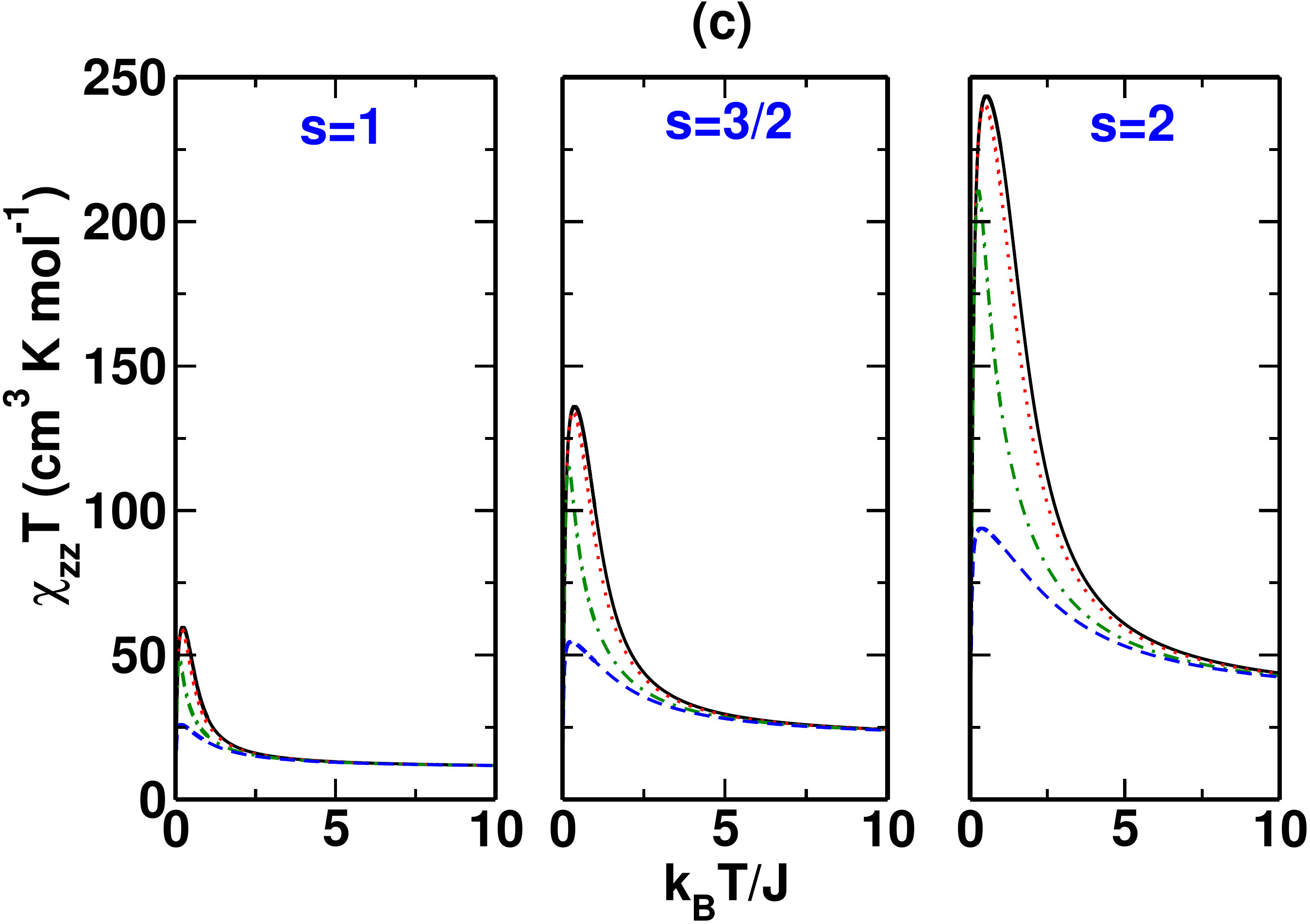} 
      \caption{\label{fig:ChiXX,chiYY,ChiZZ}(a) Plot of $\chi_{_{xx}}T$ as a function of temperature computed with applied field along x-direction ($H_x=J/g\beta=0.005$), (b) Plot of $\chi_{_{yy}}T$ as a function of temperature computed with applied field along y-direction ($H_y=J/g\beta =0.005$) and (c) Plot of $\chi_{_{zz}}T$ as a function of temperature computed with applied field along z-direction ($H_z=J/g\beta=0.005$) for different values of $\epsilon$ and $\delta$, in the absence of on-site anisotropy. Color coding and line type is the same for all panels. (Note scale for (c) are different from those of (a) and (b)).}
\end{figure*}

\begin{figure*}
\center
    \includegraphics[width=8cm]{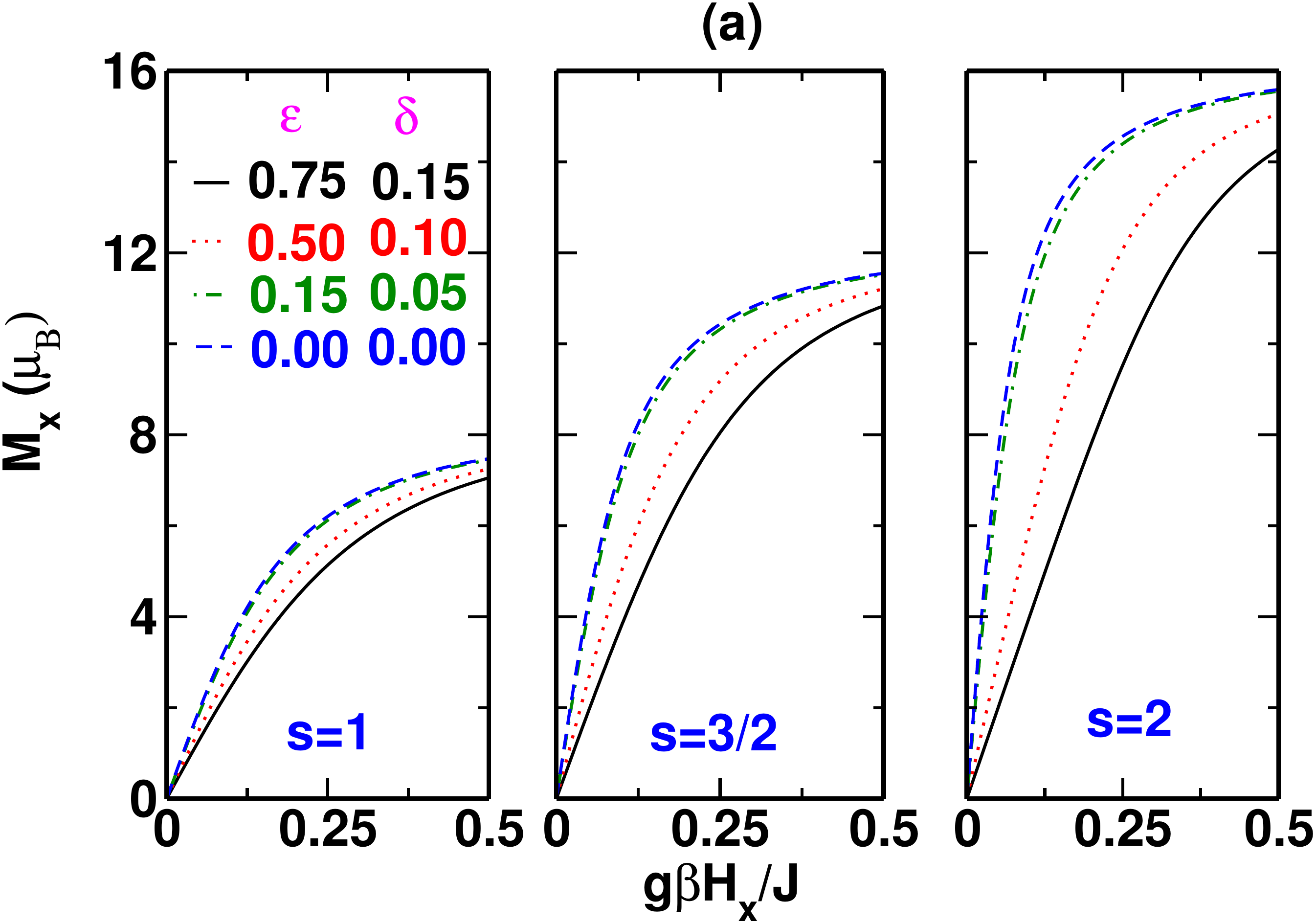} 
\center
    \includegraphics[width=8cm]{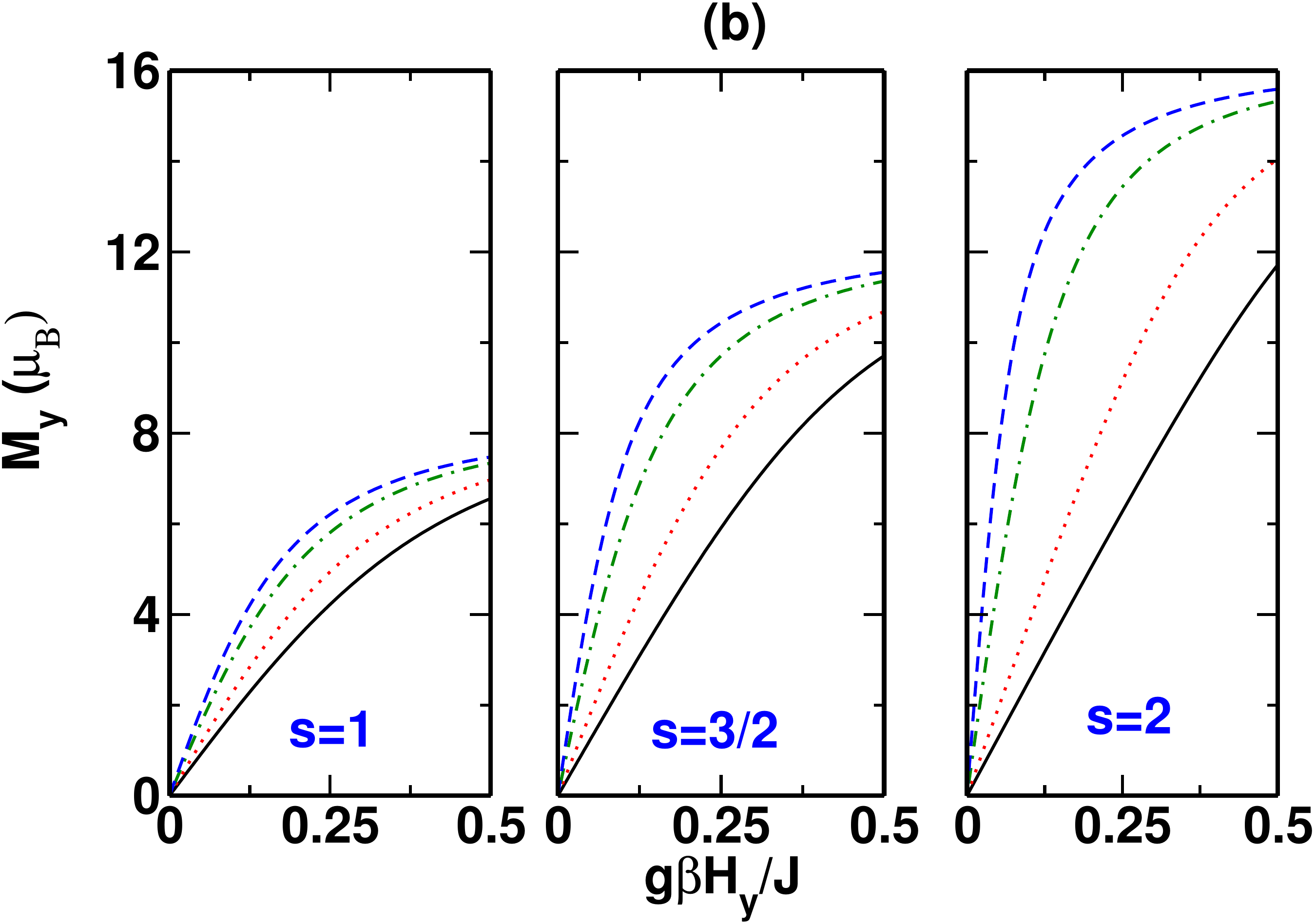}  
\center
      \includegraphics[width=8cm]{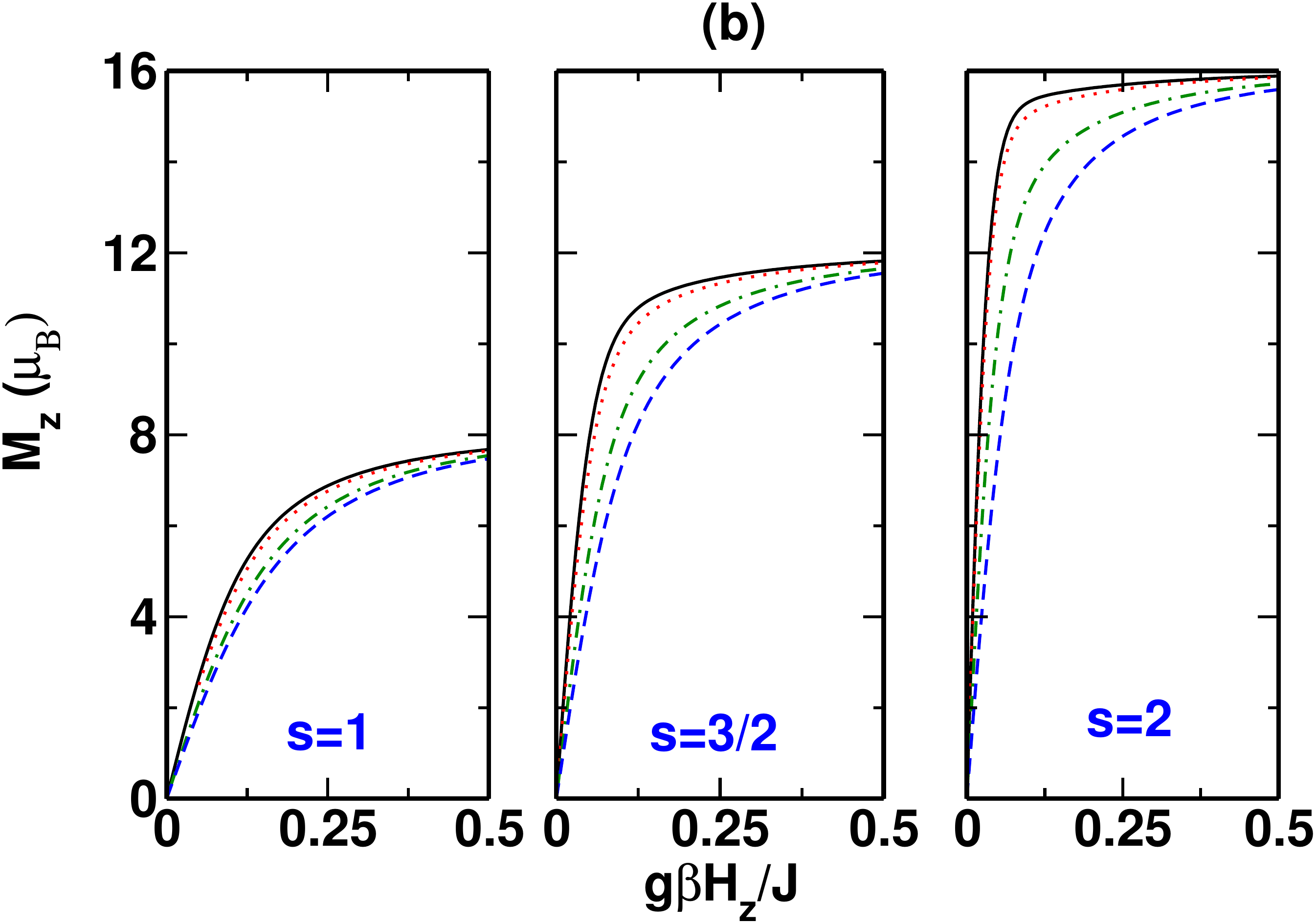} 
      \caption{\label{fig:MxHx,MyHy,MzHz} Dependence of magnetization ($M$) on applied magnetic field ($g\beta H/J$) at temperature $k_BT/J=1.0$. (a) $M_x$ versus $g\beta H_x/J$, (b) $M_y$ versus $g\beta H_y/J$ and (c) $M_z$ versus $g\beta H_z/J$ for different values of $\epsilon$ and $\delta$, in the absence of on-site anisotropy. Color coding and line type is the same for all panels.}
\end{figure*}

\begin{figure*}
    \includegraphics[width=12cm]{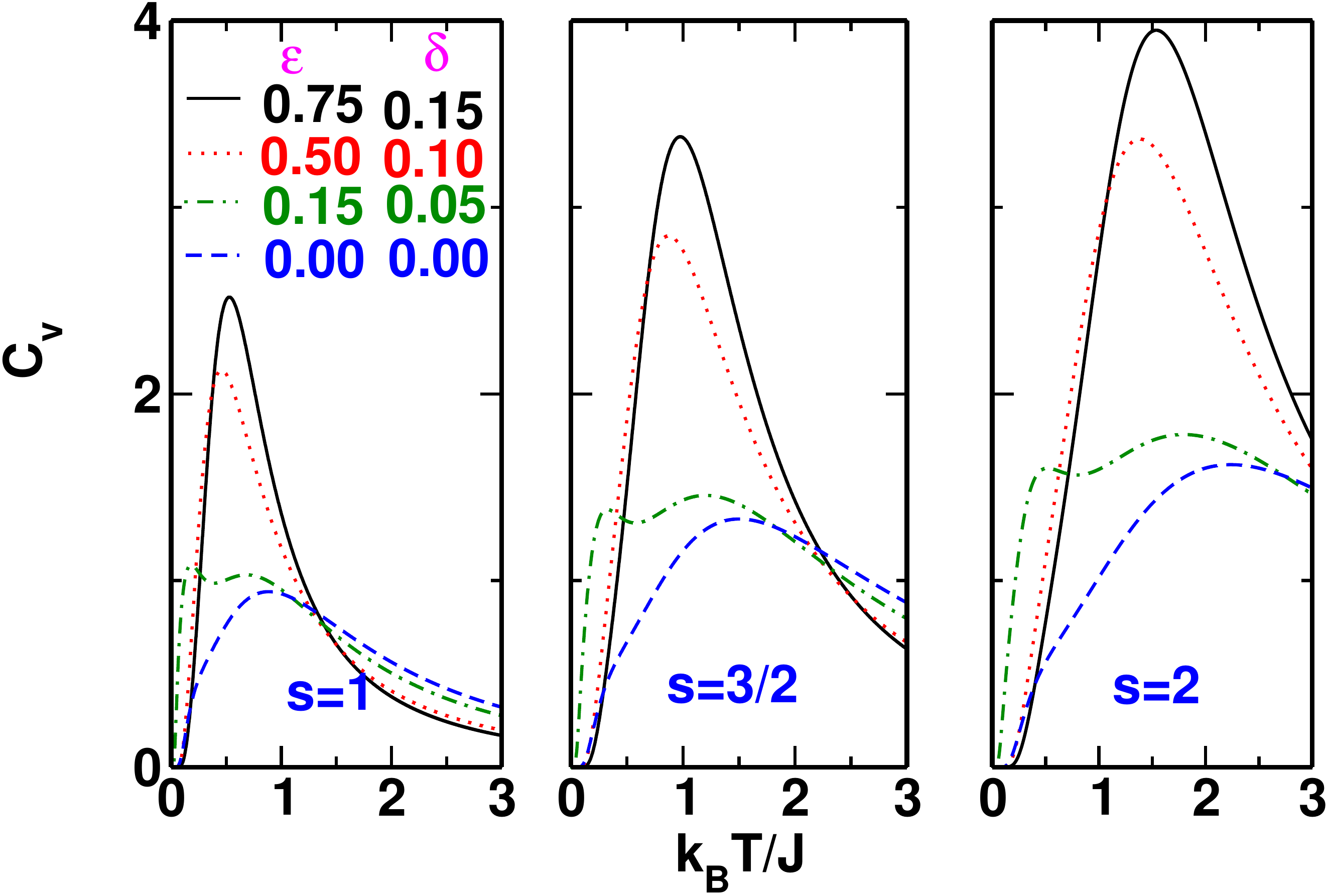} 
      \caption{\label{fig:Cvxyz}Dependence of specific heat ($C_v$) on temperature ($k_BT/J$) of spin chains with $s=1$, $3/2$ and $s=2$ for different values of $\epsilon$ and $\delta$, in the absence of on-site anisotropy. Color coding for $2^{nd}$ and $3^{rd}$ panel are same as for the first panel.}
\end{figure*}

\begin{figure}
    \includegraphics[width=10cm]{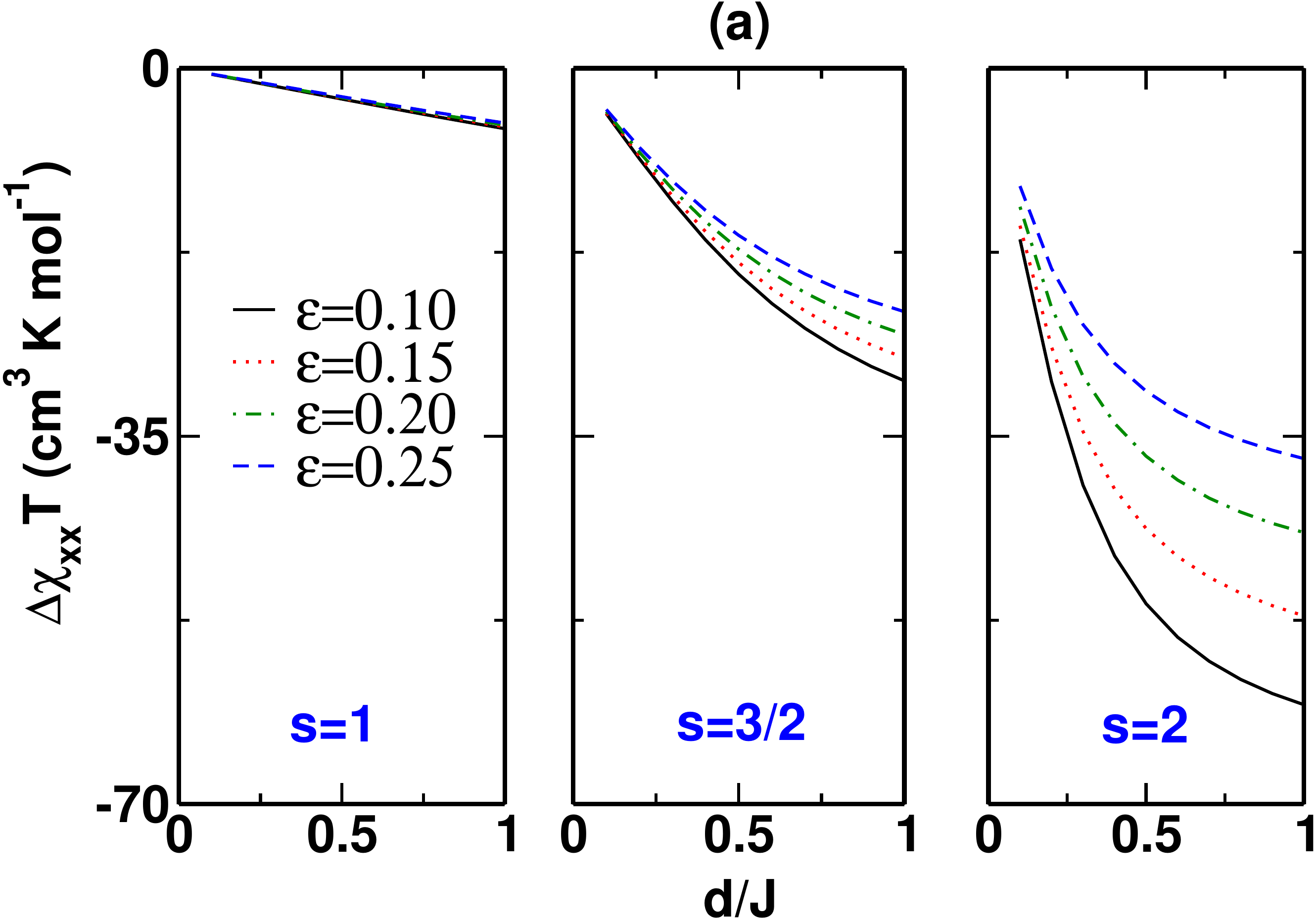}  

\vspace{0.3in}
\includegraphics[width=10cm]{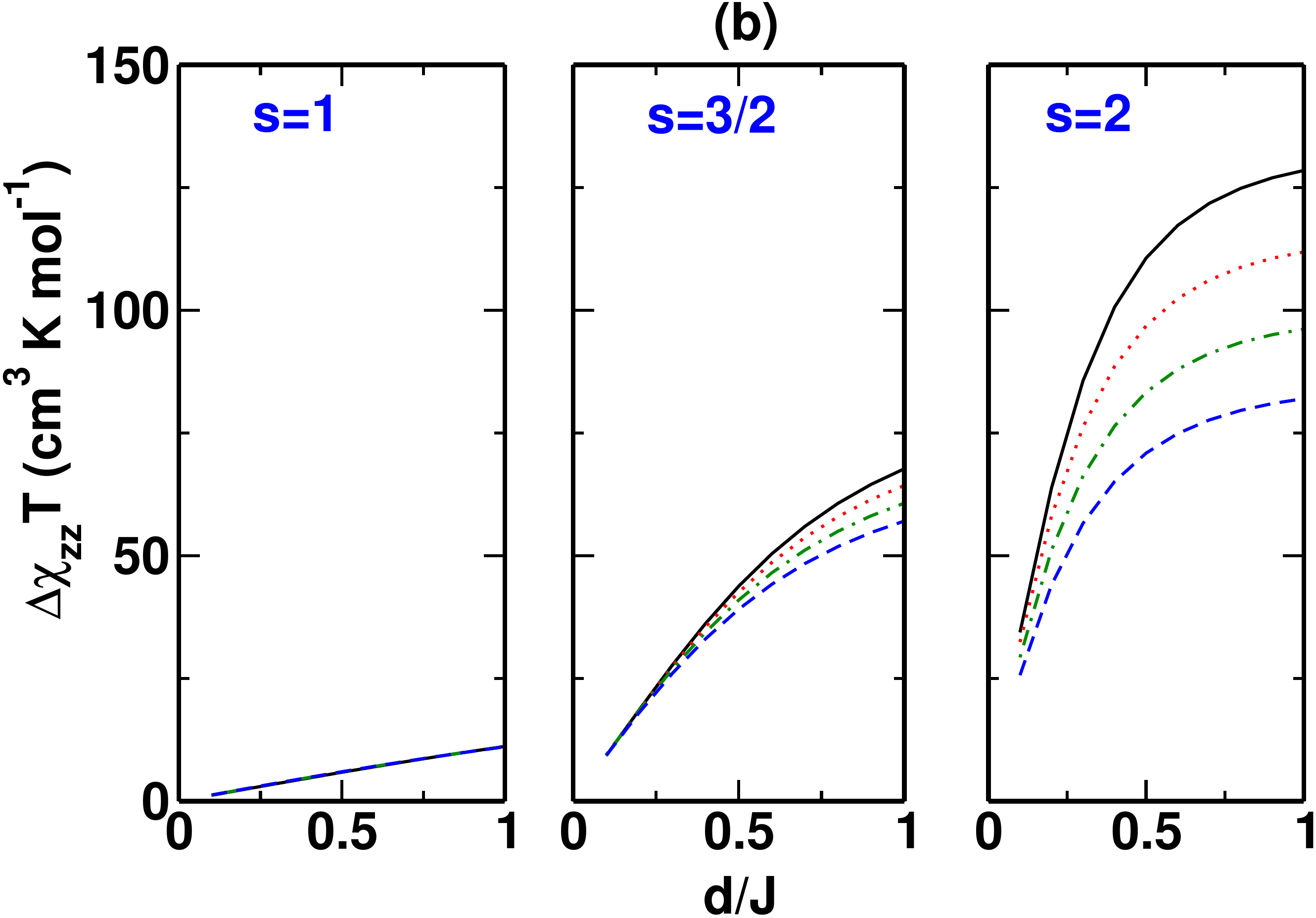} 
      \caption{\label{fig:ChiXXd,ChiZZd}The effect of on-site anisotropy $d/J$ on (a)$\Delta\chi_{_{xx}}T=[\chi_{_{xx}}T(\epsilon, d \neq 0)-\chi_{_{xx}}T(\epsilon, d=0)]$ at $g\beta H_x/J=0.005$, $k_BT/J=1.0$ and (b) $\Delta\chi_{_{zz}}T=[\chi_{_{zz}}T(\epsilon, d \neq 0)-\chi_{_{zz}}T(\epsilon, d=0)]$ at $g\beta H_z/J=0.005$, $k_BT/J=1.0$ for $\epsilon=0.10$, $0.15$, $0.20$ and $0.25$. Same color code and line type is used for all panels. Also note the sign of $\Delta\chi_{_{xx}}T$ is -ve while $\Delta\chi_{_{zz}}T$ is +ve.  }
\end{figure}

\begin{figure}
    \includegraphics[width=10cm]{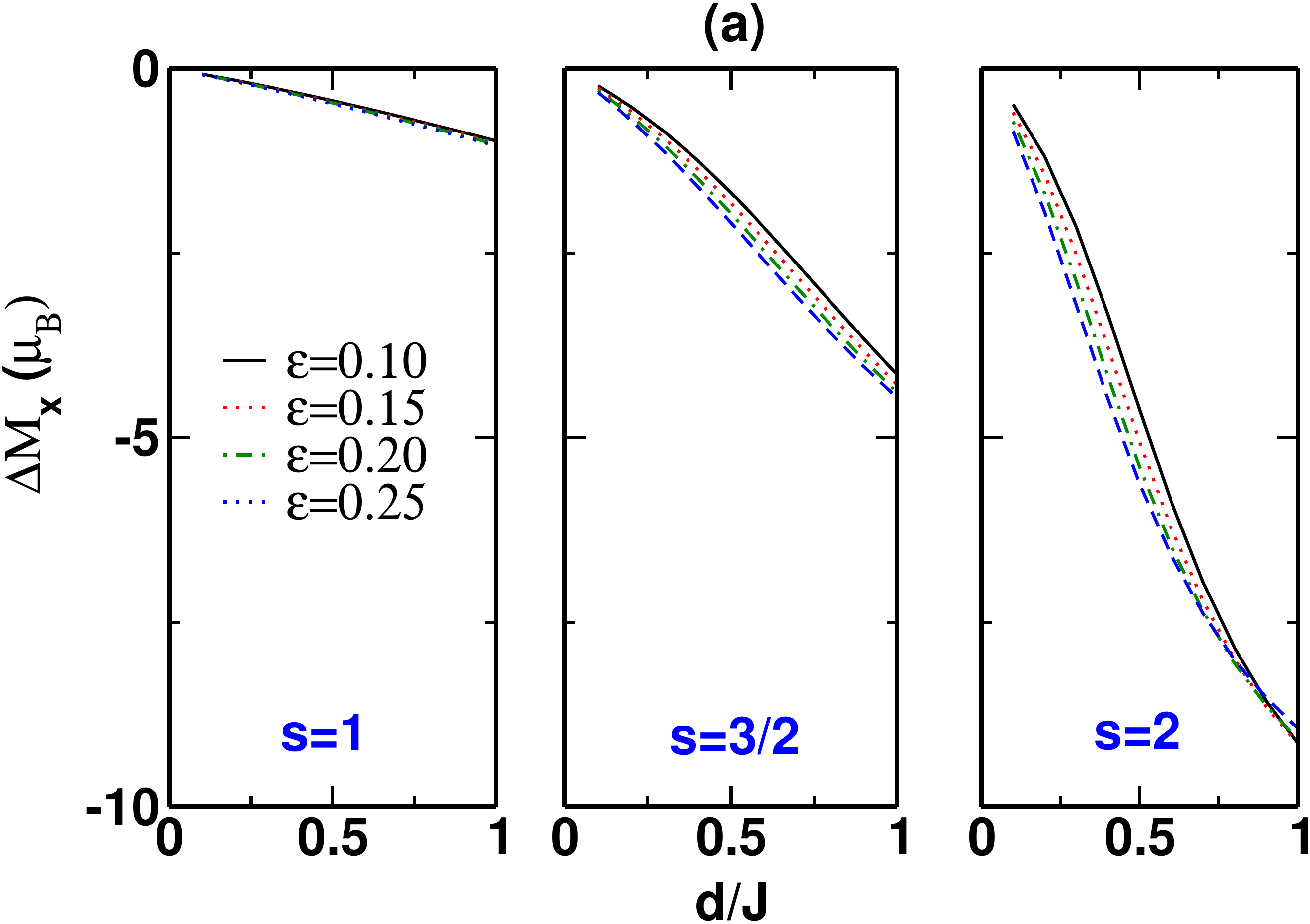}  

 \vspace{0.3in}
 \includegraphics[width=10cm]{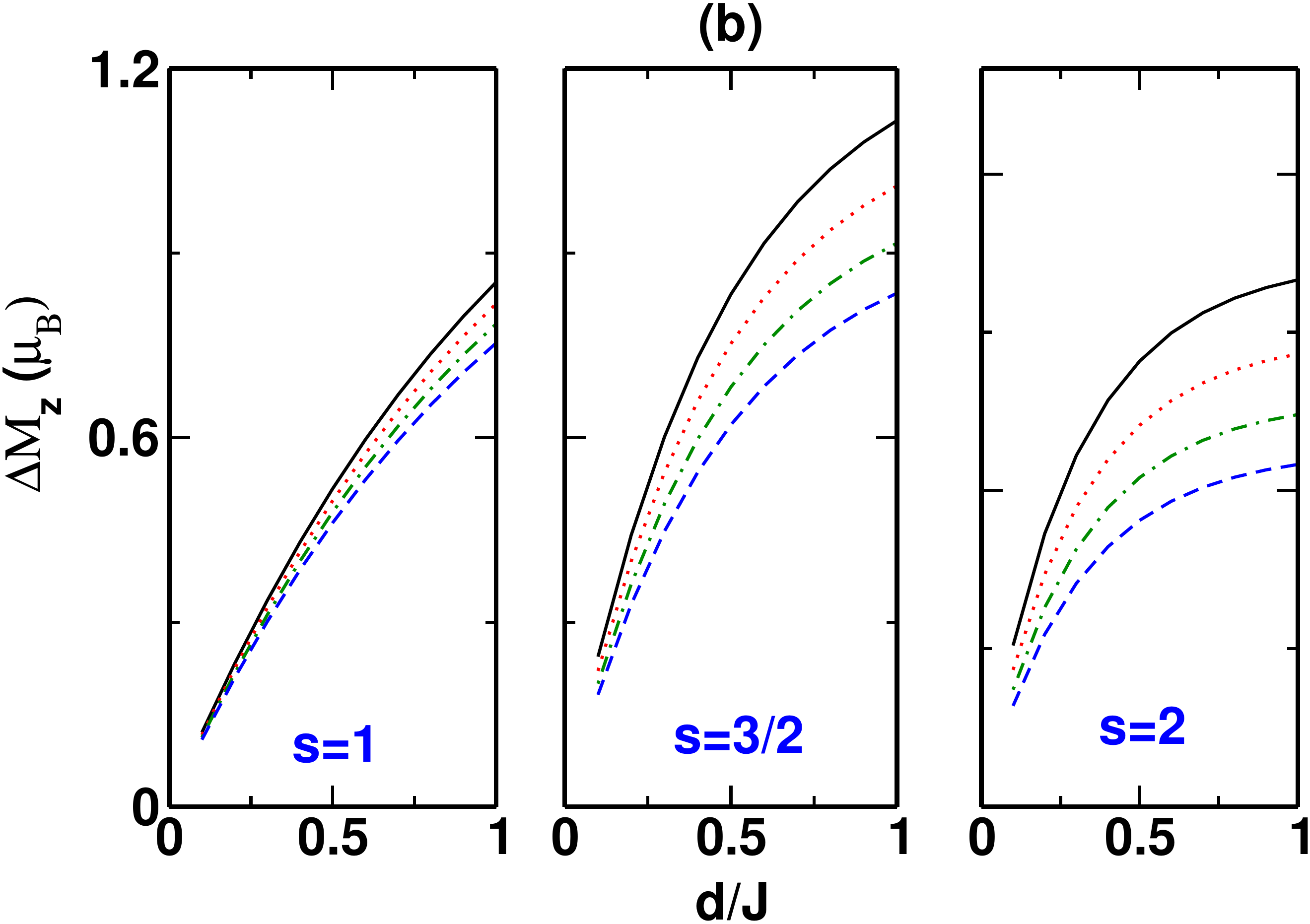} 
      \caption{\label{fig:MxHxd,MzHzd}The effect of on-site anisotropy $d/J$ on (a) $\Delta M_x=M_x(\epsilon, d \neq 0)-M_x(\epsilon, d=0)$ at $g\beta H_x/J=0.25$, $k_BT/J=1.0$ and (b) $\Delta M_z=M_z(\epsilon, d \neq 0)-M_z(\epsilon, d=0)$ at $g\beta H_z/J=0.25$, $k_BT/J=1.0$ for $\epsilon=0.10$, $0.15$, $0.20$ and $0.25$. Same color and line type is used for all panels.}
\end{figure}

\begin{figure}
\center
    \includegraphics[width=12cm]{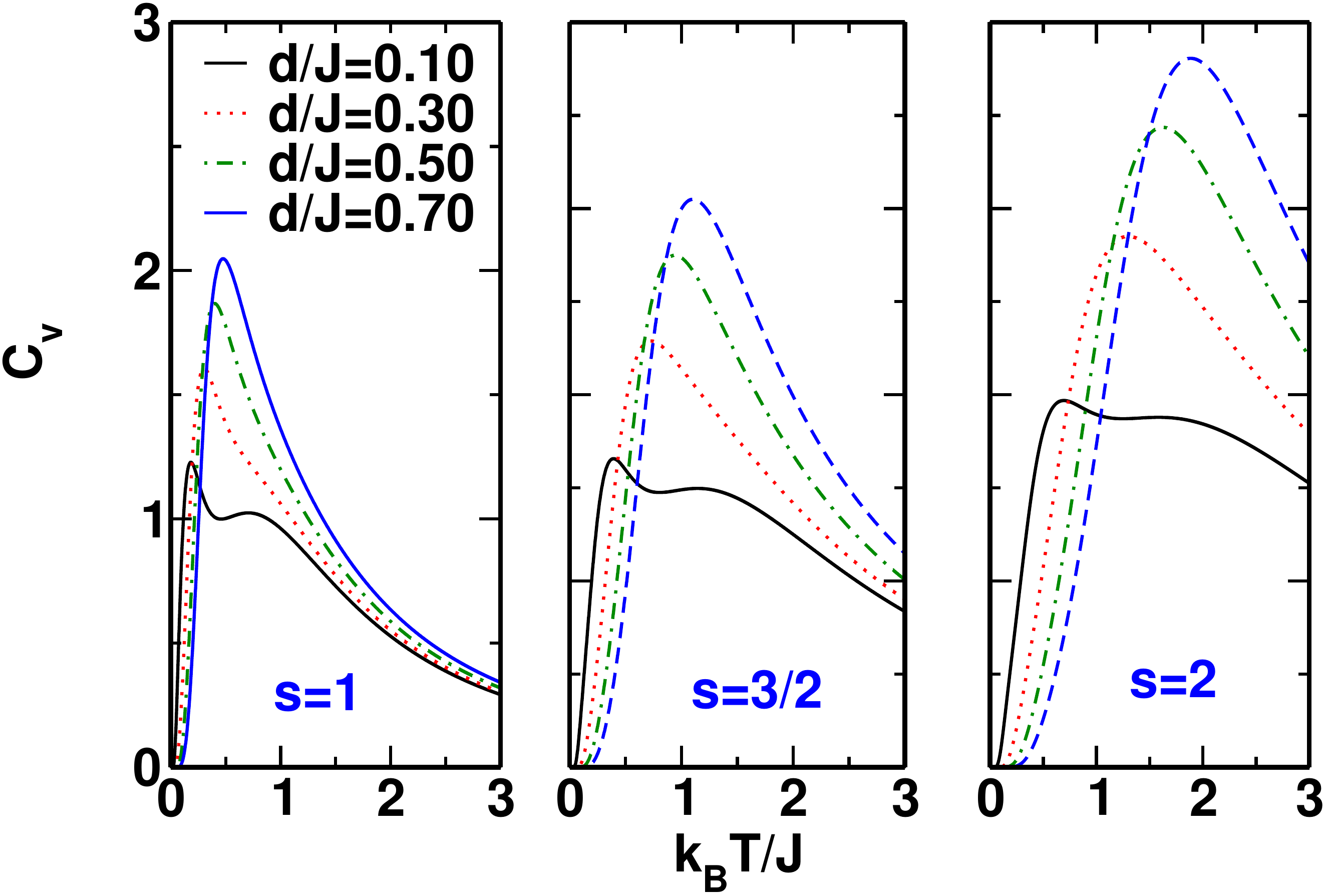} 
      \caption{\label{fig:Cvxxzd}Dependence of specific heat ($C_v$) on temperature ($k_BT/J$) of spin chains with $s=1$, $3/2$ and $s=2$ for $\epsilon=0.10$ in the presence of $d/J=0.10$, $0.30$, $0.50$ and $0.70$. Same color and line type is used for all panels.}
\end{figure}